%% file: elsarticle_main.tex
\documentclass[3p,12pt]{elsarticle}

\usepackage[utf8]{inputenc}

\usepackage[dvipsnames]{xcolor}
\usepackage{mathtools}
\usepackage{calligra}
\DeclareMathAlphabet{\mathcalligra}{T1}{calligra}{m}{n} \DeclareFontShape{T1}{calligra}{m}{n}{<->s*[2.2]callig15}{}
\usepackage{amsmath}
\usepackage{amsfonts}
\usepackage{mathrsfs}
\usepackage{graphicx}
\usepackage{relsize}
\usepackage{float}
\usepackage{svg}
\usepackage{pdflscape}
\usepackage{subfigure}
\usepackage{pdfpages}
\usepackage{comment}
\usepackage{caption}
\usepackage{textcomp}
\usepackage{gensymb}
\usepackage{placeins}

\newcommand{\bs}[1]{\boldsymbol{#1}}

\begin{document}

\title{A Non-intrusive Approach for Physics-constrained Learning with Application to Fuel Cell Modeling}
\author[AeroUM]{Vishal Srivastava}
\ead{vsriv@umich.edu}
\author[MechUM]{Valentin Sulzer}
\ead{vsulzer@umich.edu}
\author[MechUM]{Peyman Mohtat}
\ead{pmohtat@umich.edu}
\author[MechUM]{Jason B. Siegel}
\ead{siegeljb@umich.edu}
\author[AeroUM]{Karthik Duraisamy}
\ead{kdur@umich.edu}
\address[AeroUM]{Aerospace Engineering, University of Michigan, Ann Arbor, MI, USA}
\address[MechUM]{Mechanical Engineering, University of Michigan, Ann Arbor, MI, USA}

\begin{abstract}
\input{0_Abstract/main}
\end{abstract}

\maketitle

\input{1_Introduction/main}
\input{2_Background/main}
\input{3_Methodology/main}
\input{4_Results/main}
\input{5_Conclusions/main}

\bibliographystyle{elsarticle-num}
\bibliography{FC_refs}

\begin{appendix}
\input{A_SourcesAndBoundaryConditions/main}
\end{appendix}

\end{document}

%% file: 0_Abstract/main.tex
A  data-driven model augmentation framework, referred to as Weakly-coupled Integrated Inference and Machine Learning (IIML), is presented to improve the predictive accuracy of physical models.
In contrast to {\em parameter} calibration, this work seeks corrections to the {\em structure} of the model by a) inferring augmentation fields that are consistent with the underlying model, and b) transforming these fields into corrective model forms.
The proposed approach couples the inference and learning steps in a weak sense via an alternating optimization approach.
This coupling ensures that the augmentation fields remain learnable and maintain consistent functional relationships with local modeled quantities across the training dataset.
An iterative solution procedure is presented in this paper, removing the need to embed the augmentation function during the inference process.
This framework is used to infer an augmentation introduced within a Polymer electrolyte membrane fuel cell (PEMFC) model using a small amount of training data (from only 14 training cases.)
These training cases belong to a dataset consisting of high-fidelity simulation data obtained from a high-fidelity model of a first generation Toyota Mirai.
All cases in this dataset are characterized by different inflow and outflow conditions on the same geometry.
When tested on 1224 different configurations, the inferred augmentation significantly improves the predictive accuracy for a wide range of physical conditions.
Predictions and available data for the current density distribution are also compared to demonstrate the predictive capability of the model for quantities of interest which were not involved in the inference process.
The results demonstrate that the weakly-coupled IIML framework offers sophisticated and robust model augmentation capabilities without requiring extensive changes to the numerical solver. 

%% file: 1_Introduction/main.tex
\section{Introduction}

Digital transformation of industrial design and operations requires efficient reduced-fidelity models of the underlying physical phenomena.
However, in complex systems, such models contain structural inadequacies which may prevent them from providing sufficiently accurate predictions.
In the past decade, several data-driven model augmentation frameworks have been developed that aim to address such model-form inadequacies by inferring functional corrections into the baseline model from available high-fidelity data.
As an example, several such techniques have been introduced in the context of turbulence modeling which include but are not limited to genetic algorithms by Weatheritt and Sandberg \cite{Weatheritt2016}, sparse symbolic regression by Schmelzer et al \cite{Schmelzer2020}, Tensor Basis Neural Networks by Ling and Templeton \cite{TBNN2016}, Field Inversion and Machine Learning by Duraisamy et al. \cite{parish2016paradigm,singh2017machine}, Integrated Inference and Machine Learning by Holland et al \cite{FIMLC2019a, FIMLC2019b}, CFD-driven machine learning \cite{Saidi2022, Waschkowski2022}, etc.
While most of these frameworks provide promising predictive results on geometries and flow conditions similar to those seen in the training dataset, they usually require significant changes within the numerical solver and considerable expertise in both the method itself and the physical phenomena being modeled.
This work introduces a new framework that minimizes such requirements to reduce the time and effort needed to setup the inference procedure while ensuring various consistencies among the training and prediction environments.
This framework is demonstrated by augmenting a polymer electrolyte membrane fuel cell (PEMFC) model using high-fidelity data.
\bigskip

To meet the challenges of climate change and reduce automotive emissions, there has been a steady push for development of alternative power-train systems with lower emissions.
One such alternative is the hydrogen Fuel Cell (FC) \cite{Olabi2021FuelPerspective}, which is an electrochemical device that directly converts chemical energy into electricity with high efficiency.
Despite major advancements, the cost and durability of Polymer Electrolyte Membrane Fuel Cell (PEMFC) vehicles remain a challenge for their large scale adoption. For better control and management of a PEMFC, it is necessary to have computationally inexpensive physics-based models on-board a vehicle that can run in real-time with sufficient predictive accuracy \cite{Daud2017PEMReview, Yuan2020Model-basedReview}.
This is due to the fact that direct measurements of important internal states of a fuel cell are very difficult and/or prohibitively expensive in real-time \cite{Priya2018AModelling}.
For instance, one such quantity that significantly affects the performance of a PEMFC is the water content inside the membrane and gas channels.
Obtaining reliable measurements for the water content is difficult and therefore must be estimated from the model using the observed current, voltage, and temperature measurements.
There are a number of different approaches for modeling PEMFCs ranging from simple 1D models to complex 3D models \cite{Arif2020DifferentReview}.
However, reduced order models \cite{Goshtasbi2020ACells, Vetter2019FreeModel}, which meet the limited computational requirements of an embedded computer, may not achieve satisfactory performance (in terms of model accuracy) or are too difficult to calibrate due to a lack of available information on the internal system states.
\bigskip

In the past few years, machine learning methods have been used to design data-driven surrogate models and control strategies for PEMFCs, as well.
Napoli et al. \cite{Napoli2013} used classical neural networks along with stacking strategies to develop data-driven fuel cell models to predict the output voltage and cathode temperature of a fuel cell given the stack current and the flow rates for different gases.
Li et al. \cite{Li2014} used data-driven classification strategies supported by carefully chosen feature extraction and data labeling techniques for the diagnosis of water content related faults such as membrane drying and catalyst or channel flooding.
Using inlet pressures of hydrogen and oxygen, stack temperature and relative humidity as inputs, Han et al. \cite{Han2016} compared the voltage and current predictions obtained from data-driven surrogate models trained using neural networks and support vector machines.
Ma et al. \cite{Ma2018} used recurrent neural networks with G-LSTM (grid long short-term memory) neurons to train and predict the degradation to a fuel cell's performance due to impurities in the incoming hydrogen fuel or changes in the operating conditions.
Zhu et al. \cite{Zhu2019} used artificial neural networks (ANN) with considerable success to create a surrogate model for a high temperature proton exchange membrane fuel cell, which was further used to conduct a parameter study for the fuel cell geometry and operating conditions. These quantities also served as the inputs to the ANN.
Sun et al. \cite{Sun2020} used a hybrid methodology (using both model-based and data-driven) to construct optimal PID (Proportional Integral Derivative) and ADRC (Active Disturbance Rejection Control) control strategies for the fuel cell stack cooling.
Wang et al. \cite{Wang2020} used support vector machines (SVM) to create a data-driven surrogate model from 3D simulation data which was then used to optimize the catalyst layer composition using a genetic algorithm.
A common theme among the aforementioned works is that the data-driven models can predict scalar outputs like stack voltage and stack current but not field quantities within the fuel cell itself.
Secondly, most of these models are purely data-driven and do not incorporate physical laws manifested in the traditional models.\bigskip

Data-driven techniques that introduce corrections into a traditional model, instead of building a surrogate one, alleviate these issues to a large extent while also providing access to field outputs.
The simplest and earliest of such techniques includes parameter estimation, which involves optimizing a single model parameter in order to improve predictive accuracy.
Although introducing corrections in model parameter values can improve predictions to some extent \cite{Siegel2010ECS}, such an approach is unable to incorporate any additional physical correlations into the model.\bigskip

To introduce such corrections, the model form  needs to be augmented appropriately. This augmentation function has to be inferred from available high-fidelity data (i.e. data obtained from experiments or more accurate, yet computationally expensive simulations). As mentioned before, several such model augmentation frameworks exist in the literature and while these techniques have not been extensively used in the fuel cell modeling community, their predictive capability has been successfully demonstrated for problems in other disciplines, e.g. data-driven augmentation of turbulent fluid flow models.\bigskip

Even among such model augmentation techniques, the generalizability of the augmented model depends on a range of different factors including model consistency of the framework, diversity and parsimony of the training dataset, choice of functional form for the augmentation, choice of features (quantities that the augmentation is a function of), choice of technique to solve the inference problem, etc.
Since modeled quantities can behave significantly differently compared to their physical counterparts, it is important that inferred augmentation is model consistent, i.e. the augmentation is inferred as a function of the corresponding modeled quantities and not the physical quantities \cite{Duraisamy2021}.
FIML (Field Inversion and Machine Learning) was among the first frameworks to offer a versatile model-consistent framework which can be used to create augmentations with good predictive accuracy and a reasonable range of applicability.
While FIML in its original form suffers from limited learnability (see section \ref{ssec:FIML_Background}), Integrated Inference and Machine Learning (IIML) - which is based on the FIML framework - removes such shortcomings and improves the accuracy and generalizability of the augmented model.\bigskip

In this work, we develop a novel weakly-coupled IIML technique which facilitates inference of augmentation functions without embedding them into the solver.
This technique is demonstrated by applying it to improve the accuracy of an existing  fuel cell model.
The inference process attempts to minimize the discrepancy between predictions and available higher-fidelity data for the ionomer water content.
The data used here was provided by Toyota from predictions using a proprietary higher fidelity model.
It is observed that training on only a few representative cases resulted in considerable improvement in the predictive accuracy across a majority of test cases (with input parameters significantly different from those used during training).\bigskip

This paper is structured as follows.
Section \ref{sec:Background} briefly describes the low-fidelity model used for predictions followed by an introduction to different variants of the Field Inversion and Machine Learning approach available in literature.
Section \ref{sec:Methodology}, then discusses how the augmentation is introduced into the fuel cell model, how the augmented model is solved via a non-intrusive iterative method that bypasses the need to embed the augmentation within the numerical solver, and a novel weakly-coupled Integrated Inference and Machine Learning (IIML) strategy to enable a corresponding non-intrusive inference.
Training and validation results using this weakly-coupled IIML technique are then presented for the fuel cell model in section \ref{sec:Results} followed by conclusions in section \ref{sec:Conclusions}.

%% file: 2_Background/main.tex
\section{Background} \label{sec:Background}
  
  \input{2_Background/1_FuelCell/main}
  \input{2_Background/2_FIML/main}

%% file: 2_Background/1_FuelCell/main.tex
\subsection{Physical modeling of Fuel Cells} \label{ssec:FuelCellModel}
        
\begin{figure}[!h]
    \centering
    \includegraphics[width=0.5\textwidth]{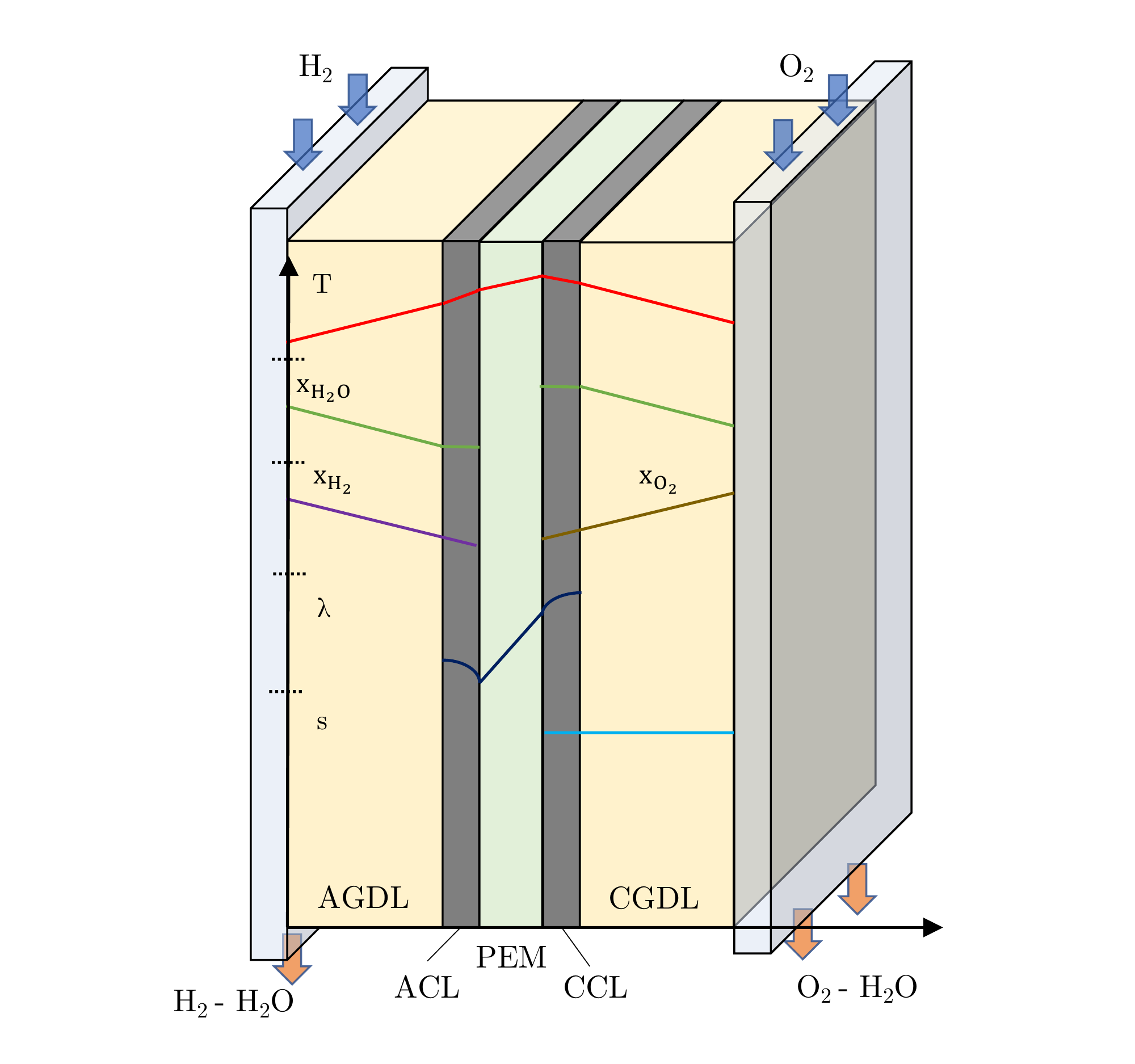}
    \caption{3D representation of a polymer electrolyte membrance fuel cell (PEMFC)}
    \label{fig:FC_schematic}
\end{figure}

\begin{figure}
    \centering
    \includegraphics[width=0.75\linewidth]{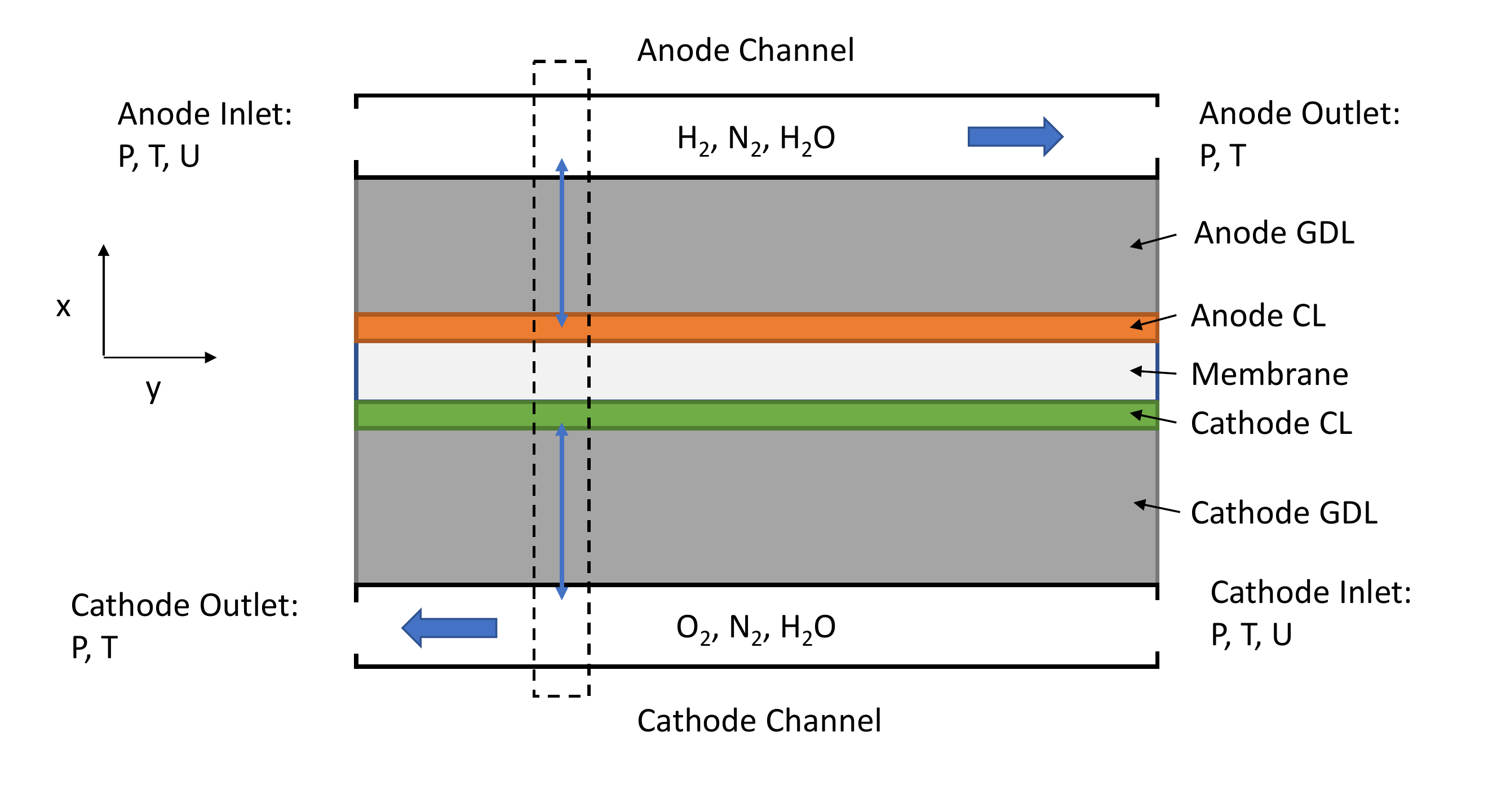}
    \caption{Schematic detailing variation of quantities within a fuel cell. The concentration and temperature gradients across the membrane, catalyst layers, and gas diffusion layers in the through-plane or x-direction are resolved by equations (1-8) at each spatial node along the y-direction. These fluxes are coupled to the along-channel or y-direction distributions by equation (9). }
    \label{fig:FuelCell}
\end{figure}

A fuel cell is an electrochemical energy conversion device that directly converts chemical energy to electrical energy.
In polymer electrolyte membrane fuel cells (PEMFC), hydrogen gas is supplied as the fuel.
A 3D representation of a PEMFC is shown in Fig. \ref{fig:FC_schematic}.
Hydrogen travels through the gas diffusion layer (GDL) to the catalyst layer.
At the anode catalyst layer, a hydrogen oxidation reaction $H_2 \longrightarrow 2H^+ + 2e^-$ produces protons and electrons.
Electrons flow through an external circuit to create an electric current, while protons cross the polymer electrolyte membrane.
Finally,  in  the  cathode  catalyst  layer,  electrons  and  protons recombine  together with oxygen/air (which is supplied to the cathode channel) to create water in an oxygen reduction reaction:
$$ \dfrac{1}{2}O_2 + 2H^+ + 2e^- \longrightarrow H_2O. $$
Modeling of fuel cells requires a description of dynamics in both the through-plane and along-channel dimensions.
A schematic is presented in Fig. \ref{fig:FuelCell} to better illustrate the structure and working of a fuel cell.
Due to the large discrepancy in length scales between these dimensions (the aspect ratio is around $10^{-3}$, with a $100\mu$m thick GDL and $10 cm$ long channels), the model is usually decomposed into a through-plane model (along the $x$-direction) an an along-the-channel model (along the $y$-direction), with coupling between the two dimensions at the GDL-channel interface only (a ‘1+1D’ model).

\subsubsection{Full through-cell model}
  
  The full through-cell model is a transient model, based on the steady-state model presented by Vetter and Schumacher \cite{Vetter2019FreeModel}.
  The modeling domains are channels, gas diffusion layers (GDLs), and catalyst layers (CLs) in the anode and cathode, with a polymer electrolyte membrane between them as shown by the dashed box in Fig.~\ref{fig:FuelCell} for transport in the x-direction.
  The subscripts $ch$, $gdl$, $cl$, and $mb$ arer used to denote a channel, gas diffusion layer (GDL), catalyst layer (CL), or polymer electrolyte membrane (PEM) domains respectively, the and superscripts $ca$ or $an$ denote the cathode and anode sides, respectively.
  The effects of the microporous layers have been neglected in this model (following \cite{Vetter2019FreeModel}).\bigskip
  
  Conservation of current and Ohm's law result in the following elliptic system relating the electron potential $\phi_e$ and the proton potential $\phi_p$, to the current densities $i_p$ and $i_e$ and the interfacial current density $j$.
  \begin{equation} \label{eqn:proton_charge_balance}
    \frac{\partial i_p}{\partial x} = aj \quad \text{where} \quad
    i_p = -\sigma_p(\lambda, T) \frac{\partial \phi_p}{\partial x}
  \end{equation}
  \begin{equation}
    \frac{\partial i_e}{\partial x} = -aj \quad \text{where} \quad
    i_e = -\sigma_e \frac{\partial \phi_p}{\partial x}.
  \end{equation}
  Here, $a$ refers to the surface area, $j$ is the reaction current density shown in the appendix, and $\sigma_p$ and $\sigma_e$ refer to the electrical conductivity for the protons and electrons, respectively.
  The conservation of the ionomer water content, $\lambda$, is enforced using the water transport model introduced by Springer \cite{Springer1991PolymerModel},
  which consists of a diffusion term and an electro-osmotic drag term, as shown in the following equation.
  \begin{equation} \label{eqn:ftr8}
    \frac{\varepsilon_i}{V_m}\frac{\partial \lambda}{\partial t} = -\frac{\partial N_\lambda}{\partial x} + S_{ad} + r_{\text{H}_\text{2}\text{O}}
    \quad \text{where} \quad
    N_\lambda = -\frac{D_\lambda (\lambda, T)}{V_m}\frac{\partial \lambda}{\partial x} + \frac{n_d (\lambda)}{F}i_p.
  \end{equation}
  Here, $\varepsilon_i$ represents the ionomer volume fraction (which is assumed constant in this model), $V_m$ refers to the equivalent volume of dry membrane, $D_\lambda$ refers to the diffusivity of the membrane, $F$ is the Faraday's constant, and $r_{\text{H}_\text{2}\text{O}}$ refers to the rate at which water is produced within the membrane as a consequence of the oxygen reduction reaction in the cathode catalyst layer.
  $S_{ad}$ is the source term which controls the adsorption/desorption of water within the ionomer membrane.
  This term is given as
  \begin{equation}\label{eqn:S_ad}
    S_{ad} = \frac{k_{ad}}{h_{cl}V_m}(\lambda_{eq}-\lambda).
  \end{equation}
  Here, $h_{cl}$ refers to the thickness of the catalyst layer, $\lambda_{eq}$ refers to the equilibrium membrane water content and is usually given as a function of temperature and relative humidity. 
  $k_{ad}$ refers to the rate of adsorption (when $\lambda<\lambda_{eq}$) or desorption (when $\lambda>\lambda_{eq}$) and is usually a function of $\lambda$ and temperature.\bigskip
  
  Gas transport is modeled using gas concentrations (denoted by $c$) instead of the typically used gas mole fractions.
  Fickian diffusion is used for the fluxes with an effective diffusivity factor to account for the reduced diffusivity in the porous medium.
  Additional source terms are used for phase changes from adsorption/desorption and evaporation/condensation.
  \begin{equation} \label{eq:augLoc}
    \frac{\partial}{\partial t}(\varepsilon_g c_{\text{H}_\text{2}\text{O}}) = -\frac{\partial N_{\text{H}_\text{2}\text{O}}}{\partial x} - S_{ad} - S_{ec}
    \quad \text{where} \quad
    N_{\text{H}_\text{2}\text{O}} = -D_{\text{H}_\text{2}\text{O}}^\text{eff}(s,T)\frac{\partial c_{\text{H}_\text{2}\text{O}}}{\partial x}.
  \end{equation}
  The gas porosity, $\varepsilon_g$, is given in terms of the liquid water saturation $s$ and the porosity $\varepsilon_p$.
  Similarly, we can obtain transport equations for hydrogen and oxygen gases, with their source terms arising from the chemical reactions.
  \begin{equation}
    \frac{\partial}{\partial t}(\varepsilon_g c_{\text{H}_\text{2}}) = -\frac{\partial N_{\text{H}_\text{2}}}{\partial x} + r_{\text{H}_\text{2}}
    \quad \text{where} \quad
    N_{\text{H}_\text{2}} = -D_{\text{H}_\text{2}}^\text{eff}(s,T)\frac{\partial c_{\text{H}_\text{2}}}{\partial x}.
  \end{equation}
  \begin{equation}
    \frac{\partial}{\partial t}(\varepsilon_g c_{\text{O}_\text{2}}) = -\frac{\partial N_{\text{O}_\text{2}}}{\partial x} + r_{\text{O}_\text{2}}
    \quad \text{where} \quad
    N_{\text{O}_\text{2}} = -D_{\text{O}_\text{2}}^\text{eff}(s,T)\frac{\partial c_{\text{O}_\text{2}}}{\partial x}.
  \end{equation}
  The liquid water saturation, $s$, is governed by the following equation.
  \begin{equation} \label{eqn:ftr46}
    \frac{1}{V_w}\frac{\partial}{\partial t}(\varepsilon_\ell c_{s}) = -\frac{\partial N_{s}}{\partial x} + S_{ec}
    \quad \text{where} \quad
    N_{s} = -\frac{D_{s}^\text{eff}(s,T)}{V_w}\frac{\partial c_{s}}{\partial x}.
  \end{equation}
  The liquid volume fraction, $\varepsilon_\ell$, is given as $\varepsilon_\ell = s\varepsilon_p$ and the capillary liquid water diffusivity, $D_s$, is given as $D_s = \dfrac{\kappa}{\mu} \dfrac{\partial p_c}{s}$.
  It should be noted that this model is isothermal, so the channel temperature is assumed uniform in the through-cell direction. The respective source term definitions can be found in \ref{app:SourceTerms}.
  
\subsubsection{1-D channel model}
  
  The 1-D through-cell model is coupled to a 1-D channel model through its boundary conditions, and the channel model governs how these boundary conditions vary along the channel spatial variable $y$.
  A counter-flow channel configuration is considered in this model as shown in Fig. \ref{fig:FuelCell}.\bigskip
  
  The anode and cathode channels have different physical channel lengths due to the design of the flow path, but must be modeled on the same 1-D grid to capture the coupling through the membrane. This mapping is achieved by considering a fixed cross-sectional area for the grid points. Thus, the spatial dimensions in each channel have been non-dimensionalized by the channel length $L_{ch}$, so that a common spatial variable $y\in[0,1]$ can be use  for computations.
  The concentrations of water, hydrogen, oxygen and nitrogen, are governed by the conservation of mass and their transport is modeled using a convective-diffusive flux.
  Thus, for any gas $k\in\lbrace\text{H}_\text{2}\text{O}, \text{O}_\text{2}, \text{H}_\text{2}, \text{N}_\text{2}\rbrace$, we have
  \begin{equation} \label{eqn:ftr13}
    \frac{\partial c_{k,ch}}{\partial t} = -\frac{1}{L_ch}\frac{\partial N_{k,ch}}{\partial y} + \frac{w}{h_{ch}}S_{k,ch}
    \quad \text{where} \quad
    N_{k,ch} = -\frac{D_{k,ch}}{L_{ch}}\frac{\partial c_{k,ch}}{\partial y} + c_{k,ch}v_{ch}.
  \end{equation}
  The gas flow velocity in the channel, $v_{ch}$, is governed by the following equation.
  \begin{equation}
    \frac{\partial v_{ch}}{\partial y} = \frac{RT_{ch}}{L_{ch}p_{ch}}\frac{w}{h_{ch}}\sum_k{S_{k,ch}}.
  \end{equation}
  The source term of a species into a channel is equal to the flux of that species from the GDL into the channel in consideration. Hence,
  \begin{equation}
  S_{k,ch}^{an} = -N_k\vert_{x=0} \quad \text{and} \quad S_{k,ch}^{ca} = N_k\vert_{x=h_{tot}}.
  \end{equation}
  To ensure the conservation of mass in the model, it is important to keep track of the liquid water in the channels.
  Any accumulated liquid water in the channel is convected away by the gas flow velocity with velocity $v_{ch}$.
  \begin{equation}
    \frac{\partial s_{ch}}{\partial t} = -\frac{1}{L_{ch}}\frac{\partial(s_{ch}v_{ch})}{\partial y} + \frac{w}{h_{ch}}S_{s,ch}.
  \end{equation}
  It is assumed that the temperature in both the channels is equal to the temperature in the cooling channel which is assumed to vary linearly in y.
  The cooling channel is oriented in the same direction as the anode channel with inlet at $y=1$ and outlet at $y=0$.
  Thus, we can write the channel temperature as
  \begin{equation} \label{eqn:ftr2}
      T_{ch} = T_{in} + \Delta T (1-y).
  \end{equation}
  Similarly, it is assumed that the pressure varies linearly in both the channels as well.
  Note that, pressure unlike temperature can be significantly different in the two channels. Thus, one may write,
  \begin{equation}
    p_{ch}^{an} = p_{in}^{an} + \Delta p^{an}(1-y) \quad \text{and} \quad p_{ch}^{ca} = p_{in}^{ca} + \Delta p^{ca}y.
  \end{equation}
  Lastly, the channel current density, $i_{ch}$, and the cathode channel potential, $\phi^{ca}_{e,ch}$, are related by Ohm's law in the channel.
  $$ i_{ch} = -\frac{\sigma_{ch}}{(L^{ca}_{ch})^2}\frac{\partial^2\phi^{ca}_{e,ch}}{\partial y^2}.$$

  Solving a full order model, with appropriately discretized through-cell and channel length scales is exceedingly computationally expensive for on-board real-time use in control systems of devices using PEMFCs.
  Thus, it is imperative to use a reduced-order model for quick computations.
  The reduced order model's inadequacies may be compensated for using data-driven techniques for model augmentation.
  This approach is demonstrated herein using integrated inference and learning on a reduced-order, asymptotic linearization of the, through-cell model by Sulzer et al. \cite{Sulzer2022} which has been coupled to the aforementioned 1-D channel model discretized with 20 node points.

%% file: 2_Background/2_FIML/main.tex
\subsection{Data-driven Model Augmentation via Field Inversion and Machine Learning} \label{ssec:FIML_Background}

Field Inversion and Machine Learning~\cite{parish2016paradigm,singh2017machine} (FIML) is a data-driven approach that helps improve the predictive accuracy of a model by inferring model inadequacies as functions of some chosen features (functions of modeled quantities).
These functions are referred to as ``augmentation'' functions.
Two main versions of the FIML framework exist in literature.
These versions, referred to as classic FIML and strongly-coupled Integrated Inference and Machine Learning (IIML) in this work, have been briefly discussed in Sections \ref{sssec:ClassicFIML} and \ref{sssec:SCIIML}, respectively.

\subsubsection{Classic FIML} \label{sssec:ClassicFIML}

  Given a model
  \begin{equation}
    \mathscr{R}_m(\widetilde{\bs{u}}_m; \bs{\xi}) = 0,
  \end{equation}
  where $\widetilde{\bs{u}}_m$ are the model states and $\bs{\xi}$ specifies the configuration (geometry, boundary conditions, etc.), a spatial field of model inadequacies, $\delta(\bs{x})$, can be appropriately introduced within the model formulation to ``augment'' the model as
  \begin{equation}
    \mathscr{R}_m(\widetilde{\bs{u}}_m; \delta(\bs{x}), \bs{\xi}) = 0.
  \end{equation}
  This optimal values of $\delta(\bs{x})$ at all spatial locations in the discretized computational domain can then be inferred such that the available high-fidelity data $\bs{y}_d$ is matched by the predictions $\bs{y}(\widetilde{\bs{u}}_m;\bs{\xi})$ as closely as possible.
  Formulating a cost function $\mathcal{C}(\bs{y}_d,\bs{y}(\widetilde{\bs{u}}_m;\bs{\xi}))$ then transforms the inference problem (``Field Inversion'') into an optimization problem as 
  \begin{equation}\begin{split}
      \delta(\bs{x}) & = \text{arg}\min_{\delta'(\bs{x})}\text{ } \mathcal{C}(\bs{y}_d,\bs{y}(\widetilde{\bs{u}}_m; \bs{\xi})) + \mathcal{T}(\delta'(\bs{x});\bs{\xi}) \\
      & \text{where }\mathscr{R}_m(\widetilde{\bs{u}}_m;\delta'(\bs{x}),\bs{\xi})=0.
  \end{split}\end{equation}
  Features $\bs{\eta}(\widetilde{\bs{u}}_m,\bs{\xi})$ (modeled quantities that the model inadequacy is assumed to be a function of) are then chosen.
  In addition, a functional form for the model inadequacy is also fixed as $\beta(\bs{\eta}(\widetilde{\bs{u}}_m,\bs{\xi});\bs{w})$, where $\bs{w}$ are the parameters of the \textit{augmentation} function, $\beta$.
  Finally, a Machine Learning technique is used to obtain the optimal parameters $\bs{w}$ such that the optimal inadequacy fields $\delta(\bs{x})$ obtained from different physical configurations are matched by the corresponding augmentation function predictions as closely as possible.
  Formulating a loss function $\mathcal{L}(\beta(\bs{\eta}(\widetilde{\bs{u}}_m,\bs{\xi});\bs{w}),\delta)$ then transforms the machine learning problem into an optimization problem as follows
  \begin{equation}\begin{split}
      \bs{w} & = \text{arg}\min_{\bs{w}'}\text{ } \mathcal{L}(\beta(\bs{\eta}(\widetilde{\bs{u}}_m,\bs{\xi});\bs{w}')\delta).
  \end{split}\end{equation}
  Finally, embedding the augmentation function within the model for predictive use, the augmented model can be given as
  \begin{equation}
    \mathscr{R}_m(\widetilde{\bs{u}}_m; \beta(\bs{\eta}(\widetilde{\bs{u}}_m,\bs{\xi});\bs{w}), \bs{\xi}) = 0.
  \end{equation}
  
\subsubsection{Strongly-coupled Integrated Inference and Machine Learning}
\label{sssec:SCIIML}
  
  While the classic FIML approach is effective in extracting augmentations from configurations sharing similar physics, the task becomes progressively harder as the configurations exhibit more diverse physical behavior.
  This inefficiency occurs due to information loss during the machine learning step which can be attributed to two reasons.
  Firstly, the field inversion problem is ill-posed and multiple solutions for $\delta(\bs{x})$ can exist which offer similar improvements in predictive accuracy.
  Since the field inversion step has no information about the features, it does not necessarily choose the solution which is most suitably expressible as a function of the chosen features.
  Secondly, solving independent field inversion problems on different configurations can give rise to augmentation fields $\delta_j(\bs{x})$ which are correlated to the features in significantly different ways.
  Both of these inconsistencies can lead to a loss of information in the machine learning step and, hence, the so-obtained augmentation function parameters $\bs{w}$ can be sub-optimal.\bigskip
  
  An integrated inference and machine learning technique (first proposed by Holland et al. \cite{FIMLC2019a, FIMLC2019b}) can address these limitations.
  In this framework, the previously separate field inversion and machine learning steps are combined into a single inverse problem that seeks to directly infer the augmentation function parameters from the available high-fidelity data.
  To achieve this, the functional form of the augmentation is embedded within the solver.
  As a consequence, the inference process is constrained to explore only those inadequacy fields that can be represented by the augmentation function.
  In addition, this also implicitly ensures the features are correlated to the augmentation in a consistent manner across all training cases.
  Mathematically, the corresponding problem statement can be written as
  \begin{equation}
    \begin{split}
    \bs{w} & = \text{arg} \min_{\bs{w}'} \mathlarger{\bigsqcup}_{j=1}^n\left( \mathcal{C}^j(\bs{y}_d^j, \bs{y}_m^j(\widetilde{\bs{u}}_m^j;\bs{\xi}^j)) + \lambda_j\mathcal{T}^j(\beta(\bs{\eta}(\widetilde{\bs{u}}_m^j;\bs{\xi}^j);\bs{w}')); \bs{\xi}^j)\right) \\ & s.t. \quad
    \mathscr{R}_m(\widetilde{\bs{u}}_m^j; \beta(\bs{\eta}(\widetilde{\bs{u}}_m^j;\bs{\xi}^j);\bs{w}')), \bs{\xi}^j) = 0.
    \end{split}
  \end{equation}
  Here, $\bigsqcup$ denotes an assembly operator that combines the cost and regularization functions from different configurations (indicated by index $j$) into a single combined objective function.
  The assembly operator can be as simple as a weighted sum.
  As can be seen, the inference and learning procedures in this technique are no longer distinct and the function parameters $\bs{w}$ are directly inferred by solving a single optimization problem.\bigskip
  
  To perform integrated inference and learning using the approach mentioned above, the augmentation function needs to be embedded into the numerical solver to enable accurate sensitivity evaluation.
  Embedding the augmentation involves significant changes to the solver code which may require considerable effort.
  When testing several augmentation candidates and/or working with an intricate solver, being able to work with a non-intrusive solution technique can save time, effort and resources while allowing increased flexibility, ease-of-use and portability.
  This work presents a novel weakly-coupled version of Integrated Inference and Machine Learning (see Section \ref{ssec:WCIIML}) that can offer the benefits of the aforementioned strongly-coupled IIML while bypassing the need to embed the augmentation into the numerical solver.
  This framework is demonstrated by augmenting the aforementioned fuel cell model.

%% file: 3_Methodology/main.tex
\section{Methodology} \label{sec:Methodology}

  This section outlines and explains the components of the weakly-coupled integrated inference and learning technique to obtain the augmentation function parameters $\bs{w}$ from available data without embedding the augmentation function within the solver.
  Section \ref{ssec:ML_Setup} briefly discusses how the model was augmented, what features were chosen, and what neural network architecture was used.
  Thereafter, section \ref{ssec:IterBetaSolver} explains the minimal changes needed to be made to the solver along with the iterative method used to solve the augmented equations and the use of finite differences to obtain sensitivities with respect to augmentation field, $\delta(\bs{x})$.
  Following that, section \ref{ssec:WCIIML} details the proposed weakly-coupled integrated inference and machine learning strategy.
  
  \input{3_Methodology/1_FuelCellModelAugmentation/main}
  \input{3_Methodology/2_NonIntrusiveMethod/main}
  \input{3_Methodology/3_WCIIML/main}

  %
  %

%% file: 3_Methodology/1_FuelCellModelAugmentation/main.tex
\subsection{Augmenting the Numerical Solver} \label{ssec:ML_Setup}

The reduced-order through-cell model along with the full channel model constitute a system of differential algebraic equations (DAEs) which are implemented in python using the PyBaMM library \cite{PyBaMM2021} and numerically solved within the CasADi framework \cite{CasADi2019} via the SUNDIALS~\cite{SUNDIALS} solver. 
After testing different ways to augment the model, the most promising approach seems to be modifying the algebraic model used to evaluate the equilibrium water content, $\lambda_\text{eq}$ (used to calculate $S_{ad}$ in Eqn. \ref{eq:augLoc}), by multiplying it with the augmentation function $\beta$.
The equilibrium water content is typically modeled as a function of temperature and relative humidity \cite{Springer1991PolymerModel}, but since the precise values of these quantities in the catalyst layer cannot be measured during FC operation this was a logical place to insert the augmentation.
Furthermore, the membrane water content is sensitive to the equilibrium value across various physical conditions, viz. dry/humid, low/high current density, low/high temperatures etc.
The augmented form of the source term $S_{ad}$ (see Eqn. \ref{eqn:S_ad}) is shown in Eqn. \ref{eqn:S_ad_aug}.
\begin{equation}\label{eqn:S_ad_aug}
  S_{ad}^\text{aug} = \frac{k_{ad}}{h_{cl}V_m}({\color[rgb]{0.7,0,0}\beta_\text{aug}(\bs{\eta}_\text{aug};\bs{w})}\lambda_{eq}-\lambda).
\end{equation}
Here, $\bs{\eta}_\text{aug}$ represents the features and $\bs{w}$ represents the parameters that characterize the augmentation function.
The feature set used for this application contained the following quantities.
\begin{enumerate}
  \item Mole fraction of water vapor in the anode channel (from Eqn. \ref{eqn:ftr13})
  \item Temperature inside the cathode channel (from Eqn. \ref{eqn:ftr2})
  \item Mole fraction of water vapor in the cathode channel (from Eqn. \ref{eqn:ftr13})
  \item Water content in the anode catalyst layer (from Eqn. \ref{eqn:ftr46})
  \item Water vapor concentration in the anode catalyst layer (from Eqn. \ref{eq:augLoc})
  \item Water content in the cathode catalyst layer (from Eqn. \ref{eqn:ftr46} solved in the cathode domain)
  \item Water vapor concentration in the cathode catalyst layer (from Eqn. \ref{eq:augLoc} solved in the cathode domain)
  \item Membrane water content (from Eqn. \ref{eqn:ftr8})
\end{enumerate}

%% file: 3_Methodology/2_NonIntrusiveMethod/main.tex
\subsection{A Non-Intrusive Iterative Method to Solve Augmented Models} \label{ssec:IterBetaSolver}
  
\subsubsection{Introducing an augmentation term into the model equations}

  To implement it in the solver, $\delta(\bs{x})$ can be declared to be an additional field variable in the domain such that it remains constant for a single run of the numerical solver.
  Thus, $\delta(\bs{x})$ at all spatial locations is initialized before every solver run with a set of available apriori values.
  Note that, since the only change to the solver code is creating a new array and multiplying its corresponding local values to a term in the model equations, the solver code needs to undergo minimal change.

\subsubsection{Solving the augmented model}

  Assuming that an augmentation function $\beta(\bs{\eta};\bs{w})$ is given, we need to solve the model as described in eqn. \ref{eqn:AugmentedModelConstraintForm}.
  \begin{equation} \label{eqn:AugmentedModelConstraintForm}
      \mathscr{R}(\widetilde{\bs{u}}_m; \delta(\bs{x}), \bs{\xi}) = 0 \quad s.t. \quad \delta(\bs{x}) = \beta(\bs{\eta}(\widetilde{\bs{u}}_m;\bs{\xi});\bs{w}).
  \end{equation}
  To do this without embedding the augmentation function $\beta(\bs{\eta})$ into the solver, one can solve the augmented model in an iterative manner as shown in eqn. \ref{eqn:IterBetaSolver}.
  \begin{equation} \label{eqn:IterBetaSolver}
      \mathscr{R}(\widetilde{\bs{u}}_{m,i+1}; \delta_i(\bs{x}), \bs{\xi}) = 0 \quad s.t. \quad \delta_i(\bs{x}) = \rho\delta_{i-1}(\bs{x}) + (1-\rho)\beta(\bs{\eta}(\widetilde{\bs{u}}_{m,i};\xi);\bs{w}).
  \end{equation}
  Here, $\rho$ is a relaxation factor to avoid stability issues in the numerical solver.
  $\delta^{(0)}(\bs{x})$ can assume a constant value of $0$ or $1$ throughout the domain depending on whether the augmentation term is additive or multiplicative, respectively.
  An augmentation residual can be defined as
  \begin{equation} \label{eqn:AugResid}
    R_\text{aug}=\left\lVert\delta_i(\bs{x})-\delta_{i-1}(\bs{x})\right\rVert_2.
  \end{equation}
  A stopping criterion of $R_\text{aug}<10^{-3}$ worked well for the simulations performed in this work.
  While convergence is not guaranteed, an overwhelming number of the configurations tested in this work converged, while the remaining exhibited an oscillatory behavior in the augmentation residual.
  It is noteworthy here, that since the augmentation field changes in increasingly smaller amounts from one augmentation iteration to the next (given an appropriate value of the relaxation factor $\rho$), the computational cost required for the solver to converge keeps decreasing as iterations progress.
  Thus, carefully choosing the convergence criterion can be instrumental in significantly reducing the computational costs associated with the aforementioned iterative solution method.

%% file: 3_Methodology/3_WCIIML/main.tex
\subsection{Weakly-coupled Integrated Inference and Machine Learning} \label{ssec:WCIIML}

\begin{figure}[!h]
    \centering
    \includegraphics[width=\textwidth]{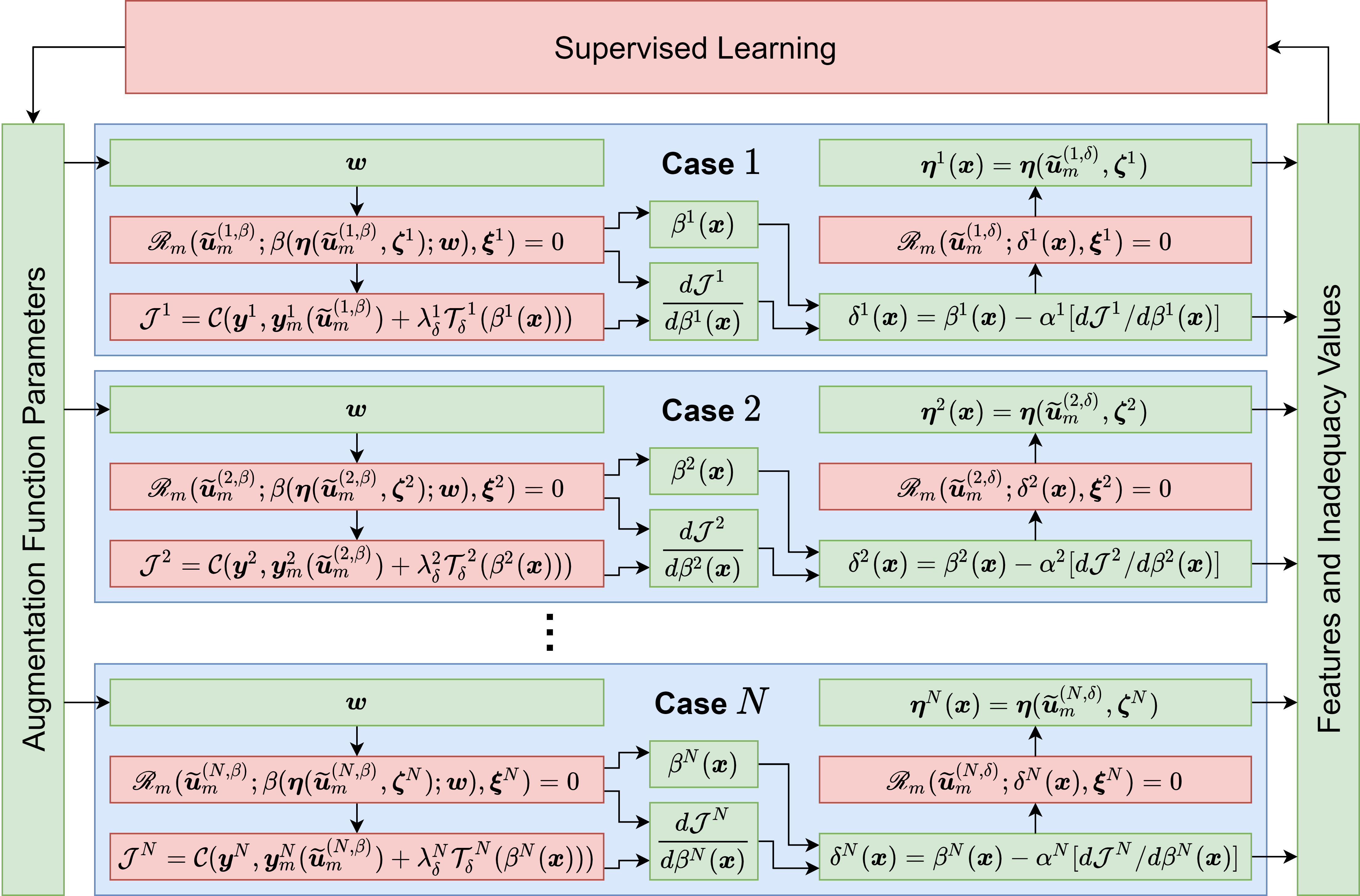}
    \caption{Schematic of weakly-coupled IIML procedure}
    \label{fig:WCIIMLflowchart}
\end{figure}

This version of IIML constrains the inadequacy field to stay consistent with the functional form chosen for the augmentation by solving the field inversion and machine learning problems in a predictor-corrector fashion while simultaneously inferring from multiple data sources.
This is done by learning the augmentation each time the inadequacy field is updated, i.e. after every iteration of field inversion.
Note that while the inadequacy fields are updated independently for all training cases, the machine learning step acts a synchronizing step for these individual optimization problems.
Data from the inadequacy fields ($\delta^i(\bs{x})$) and corresponding feature fields ($\bs{\eta}^i(\bs{x})$) is collated from all training cases and a sufficient number of machine learning iterations (epochs) are performed to ensure that the feature-to-augmentation map learns any new information from the updated flow fields.                                                      
After the machine learning step, a ``field correction'' is performed by solving the model again with the newly learned augmentation function ($\beta(\bs{\eta}(\widetilde{\bs{u}}_m,\bs{\zeta});\bs{w})$).
When the simulation converges, the predicted augmentation field ($\beta^i(\bs{x}) = \beta(\bs{\eta}(\widetilde{\bs{u}}_m^{(i,\beta)},\bs{\zeta}^i);\bs{w})$) is used as the input for the next field inversion iteration.
The superscript $(i,\beta)$ denotes that the velocity field corresponds to the $i^\text{th}$ training case and is obtained by solving the model with the augmentation \textit{function} (not the inadequacy \textit{field}).
Solving the model again with the updated augmentation function is crucial to ensure that the model predictions are consistent with the augmentation function throughout the inference process.
Finally, the sensitivity $\dfrac{d\mathcal{J}}{d\beta(\bs{x})}$ is calculated and the inadequacy field is updated using a steepest gradient method, similar to a field inversion iteration.
The step length $\alpha^i$ needs to be set manually.
In this work, it was set to $\displaystyle \frac{0.05}{\left\lVert\dfrac{d\mathcal{J}^i}{d\beta^i(\bs{x})}\right\rVert_\infty}$.
In summary, the following three consistencies are ensured when using the weakly-coupled IIML described above.
\begin{enumerate}
    \item Formulating the objective as a function of model predictions ensures that the inadequacy field iterates $\beta^i(\bs{x})$ are model-consistent.
    \item Machine learning ensures that inadequacy field iterates $\beta^i(\bs{x})$ are always consistent with the functional form of the augmentation function across all iterations.
    \item Field correction ensures that the augmentation field iterates $\beta^i(\bs{x})$ is always consistent with the augmented model.
\end{enumerate}
A flowchart describing this process is shown in Fig. \ref{fig:WCIIMLflowchart}.\bigskip

It should be noted here that the optimization trajectory for weakly-coupled IIML could be significantly different from that for its strongly-coupled counterpart.
The reason for this is explained as follows.
For the $i^\text{th}$ training case (the computational domain for which consists of $N^i_x$ discrete spatial location) the discretized inadequacy field can be represented in an $\mathbb{R}^{N^i_x}$.
Now, the set $\Delta^i$ consisting of all inadequacy fields $\delta^i(\bs{x})$ for which there exist some set of parameters $\bs{w}$ such that $\delta^i(\bs{x}) = \beta(\bs{\eta}(\widetilde{\bs{u}}_m^i,\bs{\zeta}^i);w)$ and $\mathscr{R}_m(\widetilde{\bs{u}}_m^i;\delta^i(x),\bs{\xi}^i)=0$, will form a nonlinear manifold in $\mathbb{R}^{N^i_x}$.
Strongly-coupled IIML is, by structure, constrained to explore only this nonlinear manifold.
The field inversion process (which only consists of gradient-descent-based inadequacy field updates), however, is free to find an optimal solution in the entire $N_x^i$-dimensional space.
By introducing the machine learning and field correction steps between gradient-descent-based inadequacy field updates, the weakly-coupled IIML performs a nonlinear projection operation from a point in the $N_x^i$-dimensional space to a point within the learnable manifold.
Hence, within each inference iteration, the inadequacy field can jump out of the learnable manifold after the gradient-descent-based update and is projected back into the manifold by the machine learning and field correction step.
This difference in how the iterations progress for the strongly- and weakly-coupled IIML can result in different optimization trajectories within the manifold.
While the two techniques may converge to the same solution, the difference in optimization trajectories may cause the weakly-coupled IIML -- in some cases -- to converge to a different local minima compared to strongly-coupled IIML.\bigskip

In this work, the functional form for the augmentation was chosen to be a neural network with 2 hidden layers containing 7 nodes each.
The sigmoid activation function was used in the hidden layers.
The ReLU activation function was used in the output layer to ensure that the augmentation was non-negative.
The Keras library \cite{Keras} was used to create and train the network.
The Adam optimizer \cite{Adam} was used to train the model for a total of 500 epochs after every gradient-descent-based update of the augmentation field.
The learning rate was set to be $10^{-3}$.

%% file: 4_Results/main.tex
\section{Results}
\label{sec:Results}
  The available dataset contains 1224 cases, each uniquely characterized by different inflow conditions.
  The high-fidelity data used to infer the augmentation function is the corresponding steady-state x-averaged membrane water content.
  The cost function for any case with ID $j$ was defined as 
  \begin{equation}
    \mathcal{C}_j = \left\lVert\lambda_j - \lambda_{\text{data},j}\right\rVert_2^2,
  \end{equation}
  where $\lambda$ refers to the spatial field of the membrane water content along the channel direction $y$.
  Since no regularization is used, the cost function is identical to the individual objective function for a given case.
  The combined objective function for all the cases is calculated  as the weighted sum of the individual cost functions of all training cases with all weights set to unity. Mathematically,
  \begin{equation} \label{eqn:CombObj}
      \mathcal{J} = \sum_i \alpha_i \mathcal{C}_i,
  \end{equation}
  where $\alpha_j$ represents the weights for the $j^{th}$ case which in this particular instance are all set to 1.
  
  The spatial domain used to solve the model is discretized along the channel into 20 spatial nodes.
  The reduced-order model is used to obtain the steady-state solution by running it for a sufficiently long amount of physical time which in this case was 1000 seconds.
  Due to the relatively low dimensionality of the spatial discretization, finite differences were found feasible to obtain the sensitivities of the cost function w.r.t. the augmentation field, $\beta$.
  The step-size used for finite differences was $10^{-4}$.
  The model was trained on only 14 configurations out of 1224.
  The corresponding IDs for these training cases in the dataset are 40, 100, 125, 155, 190, 230, 400, 685, 740, 840, 865, 1000, 1090 and 1200.
  These cases were chosen arbitrarily to include different kinds of input conditions (e.g. different values of relative humidity, channel temperatures, cell current densities, and stoichiometric ratios)
  
\begin{figure}[!h]
      \centering
      \includegraphics[width=\textwidth]{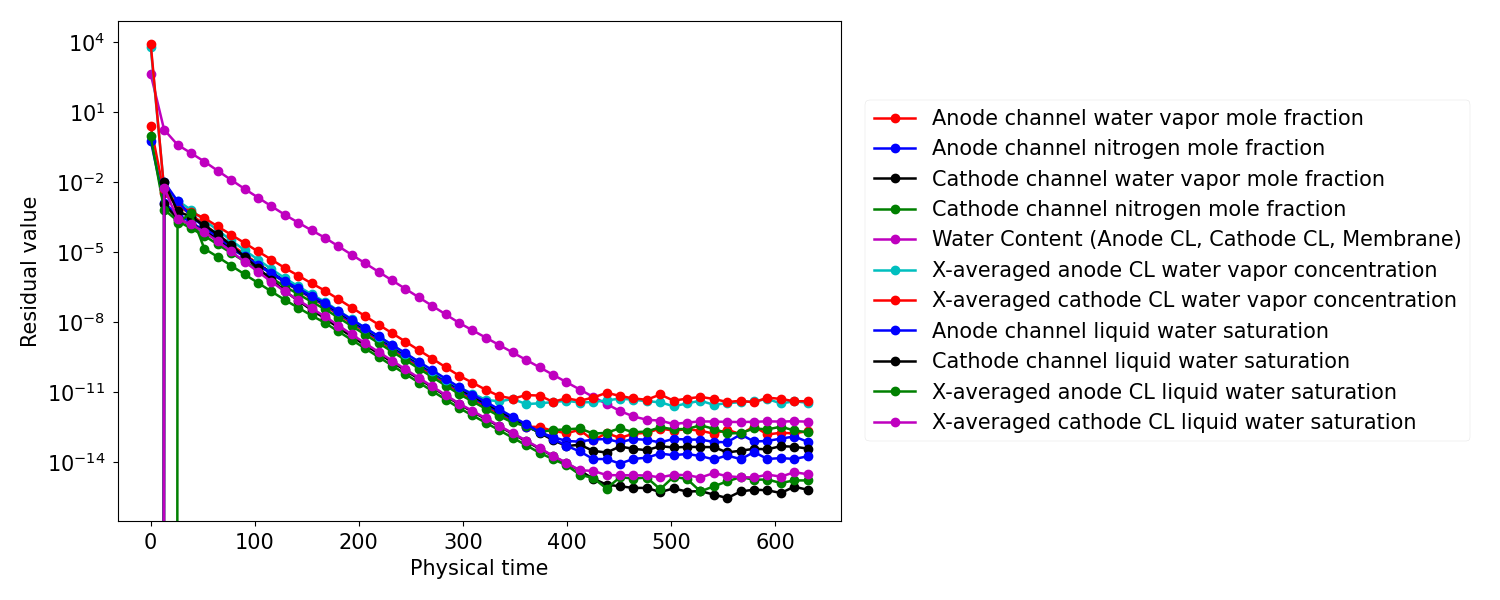}
      \caption{Representative plot for residual decay of all the state variables being solved for by the model}
      \label{fig:iterSolve}
  \end{figure}

  A representative plot for the residual histories of the states being solved for a given augmentation field is shown in Fig. \ref{fig:iterSolve}. As can be seen, the residuals approach zero within the chosen time interval that the model is solved for. It must be noted that no residual-based stopping criteria is built into the DAE solver used for this work.
  
\begin{figure}[!h]
      \centering
      \includegraphics[width=0.6\textwidth]{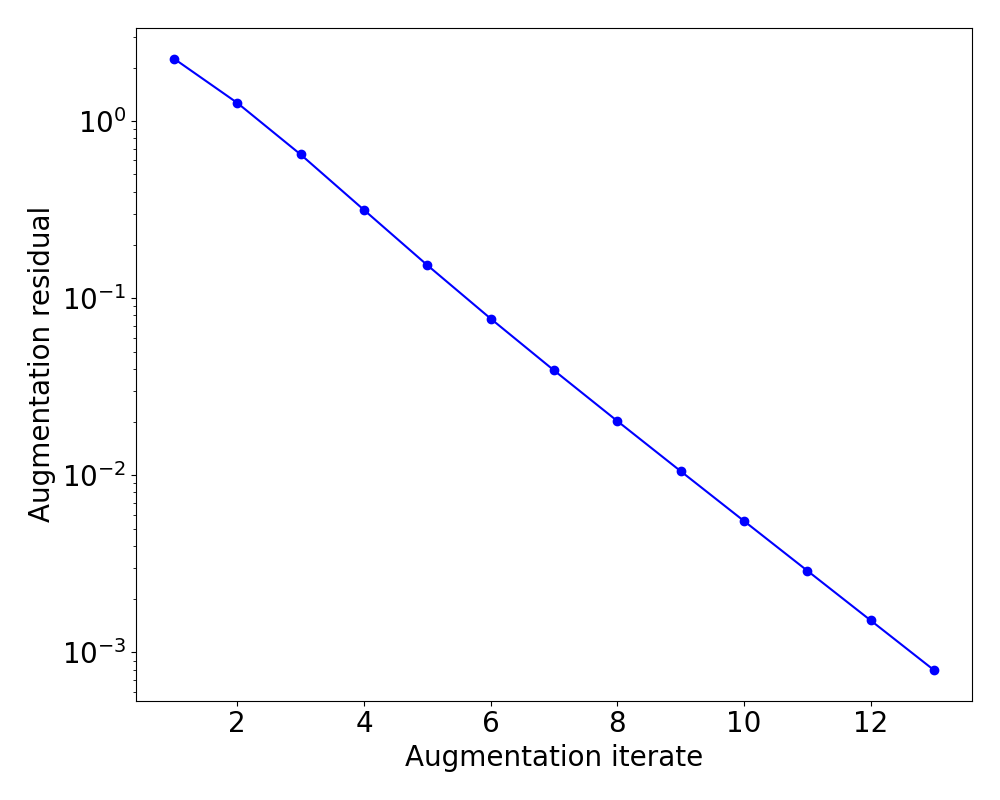}
      \caption{Representative plot for $R_\text{aug}$ convergence across the iterative solution of a model with non-embedded augmentation (from Eqn. \ref{eqn:AugResid})}
      \label{fig:iterAug}
  \end{figure}

  A representative plot for the augmentation residual ($R_\text{aug}$) history resulting from the iterative solution of the model with a non-embedded augmentation is presented in Fig. \ref{fig:iterAug}.
  Such iterative solutions need to be performed during both, training and prediction.
  Note here that a stopping condition of $R_\text{aug}<10^{-3}$ was found sufficient to obtain a sufficiently converged result.
  While there do exist a few cases where such a convergence cannot be achieved and the residuals keep oscillating, no cases exhibit divergent behavior.
  Even in the cases where the augmentation residuals keep oscillating, the residual magnitudes are very small (of the order of $10^-2$).
  
  \subsection{Training}
  
  \begin{figure}[!h]
      \centering
      \includegraphics[width=0.6\textwidth]{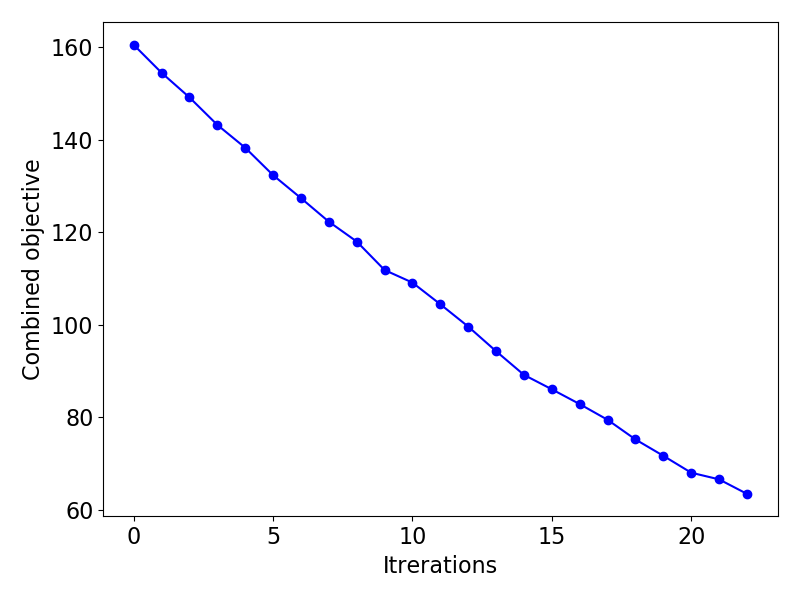}
      \caption{Minimization of objective function with inference (weakly-coupled IIML) iterations (from Eqn. \ref{eqn:CombObj})}
      \label{fig:optim}
  \end{figure}
  
  \begin{figure}[!ht]
    \centering
    \subfigure[$\mathscr{P}_1^\lambda=0.27$, $\mathscr{P}_2^\lambda=0.09$]
    {\includegraphics[width=0.31\textwidth]{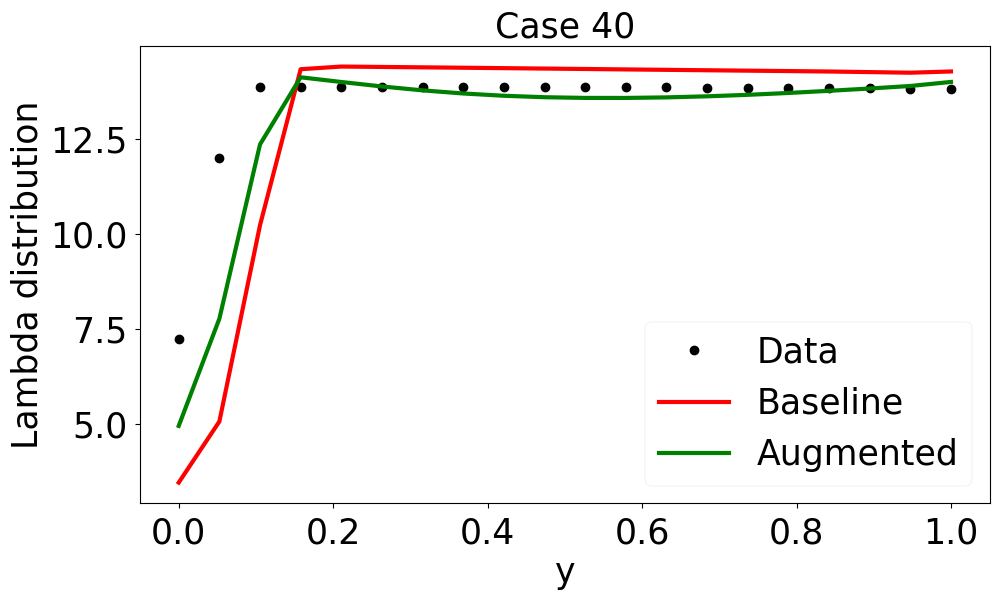}}
    \subfigure[$\mathscr{P}_1^\lambda=0.18$, $\mathscr{P}_2^\lambda=0.03$]
    {\includegraphics[width=0.31\textwidth]{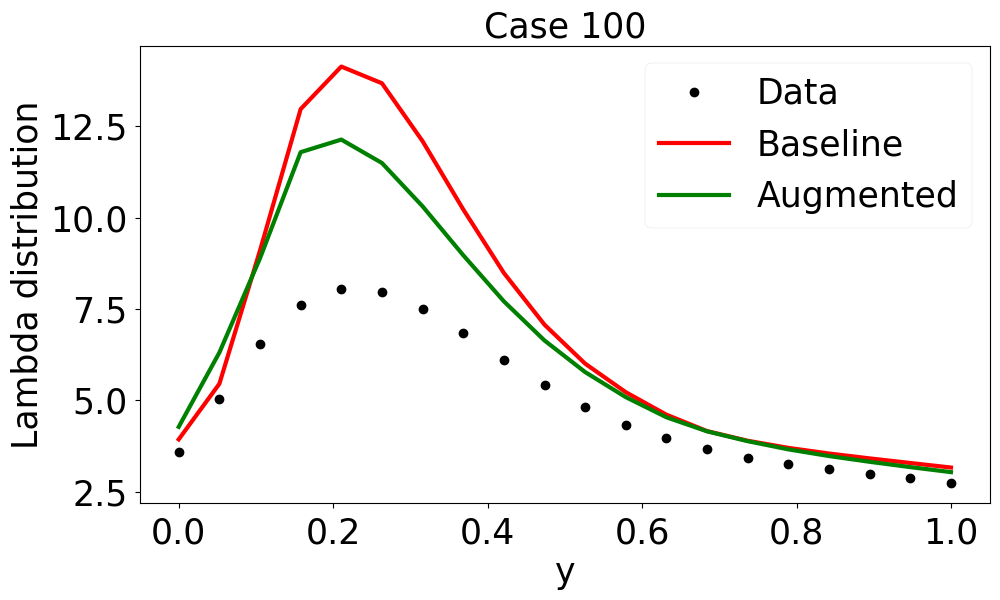}}
    \subfigure[$\mathscr{P}_1^\lambda=0.08$, $\mathscr{P}_2^\lambda=0.008$]
    {\includegraphics[width=0.31\textwidth]{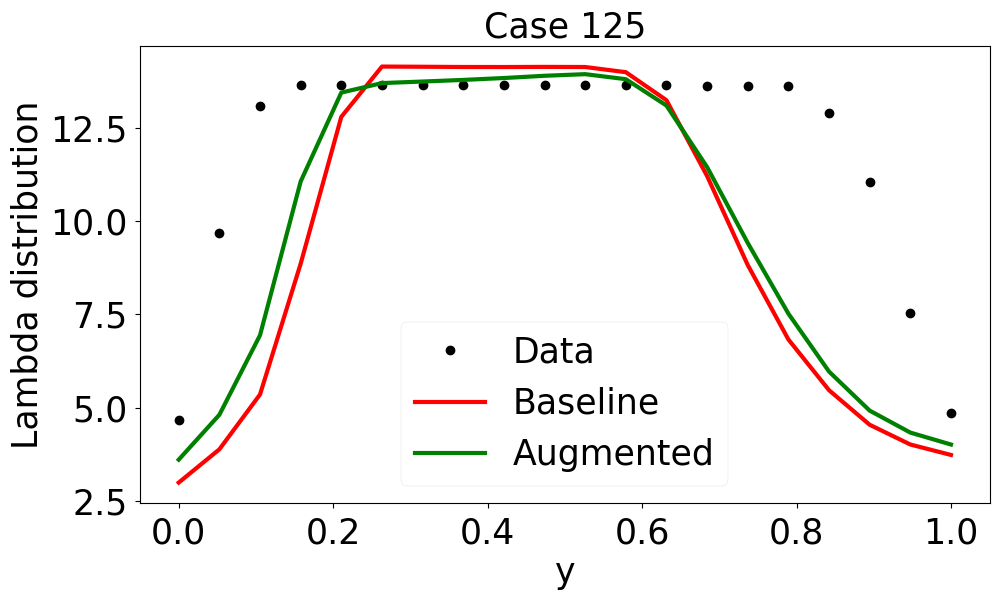}}
    \subfigure[$\mathscr{P}_1^\lambda=0.31$, $\mathscr{P}_2^\lambda=0.1$]
    {\includegraphics[width=0.31\textwidth]{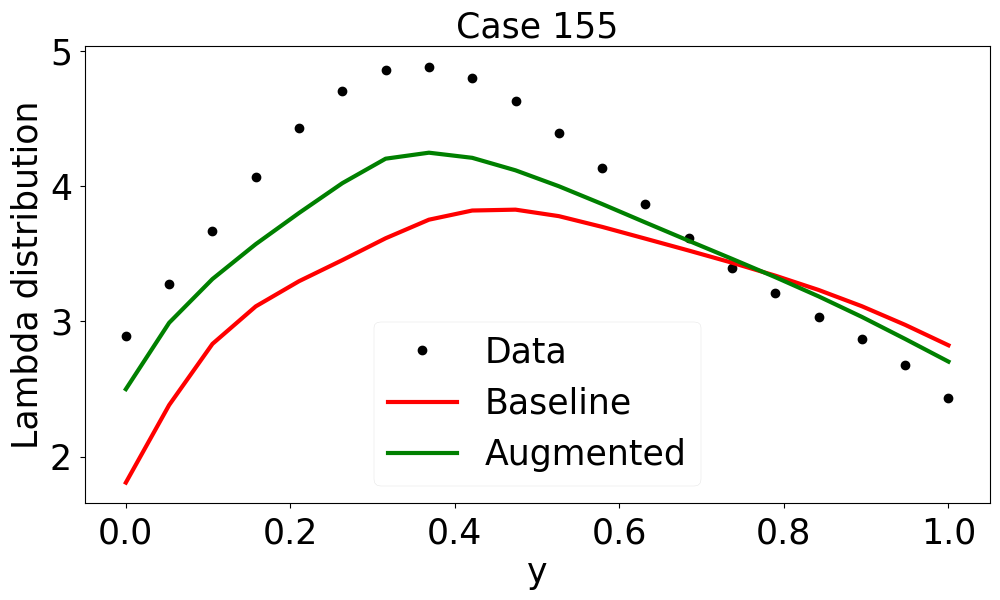}}
    \subfigure[$\mathscr{P}_1^\lambda=0.04$, $\mathscr{P}_2^\lambda=0.004$]
    {\includegraphics[width=0.31\textwidth]{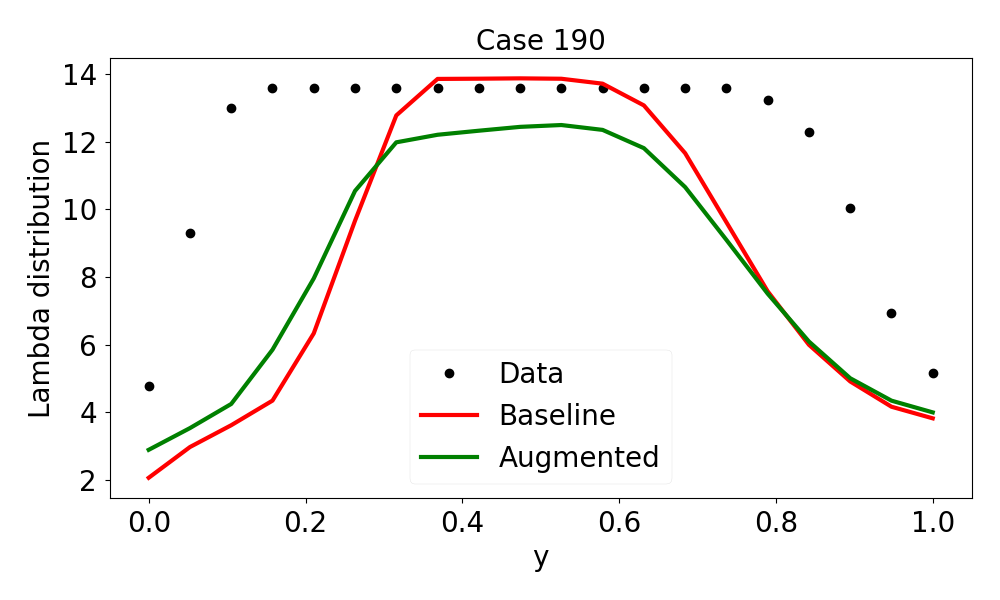}}
    \subfigure[$\mathscr{P}_1^\lambda=0.42$, $\mathscr{P}_2^\lambda=0.18$]
    {\includegraphics[width=0.31\textwidth]{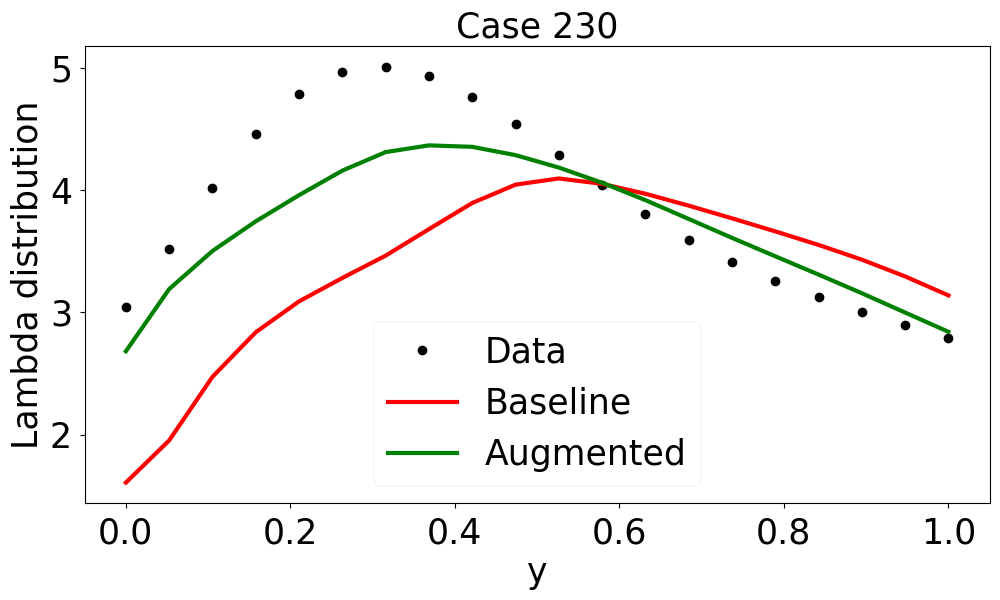}}
    \subfigure[$\mathscr{P}_1^\lambda=0.62$, $\mathscr{P}_2^\lambda=0.46$]
    {\includegraphics[width=0.31\textwidth]{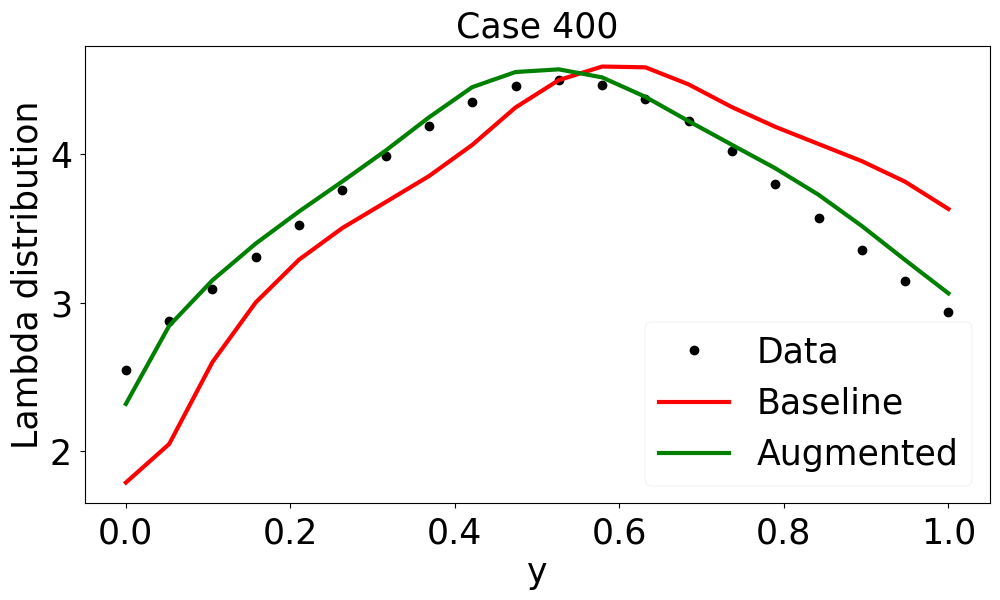}}
    \subfigure[$\mathscr{P}_1^\lambda=0.44$, $\mathscr{P}_2^\lambda=0.25$]
    {\includegraphics[width=0.31\textwidth]{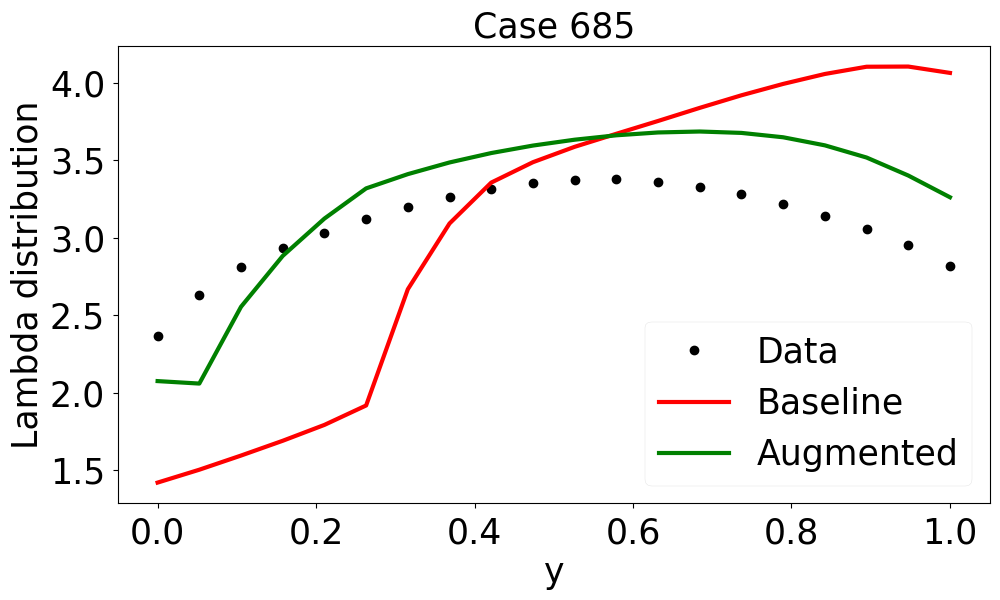}}
    \subfigure[$\mathscr{P}_1^\lambda=0.86$, $\mathscr{P}_2^\lambda=0.85$]
    {\includegraphics[width=0.31\textwidth]{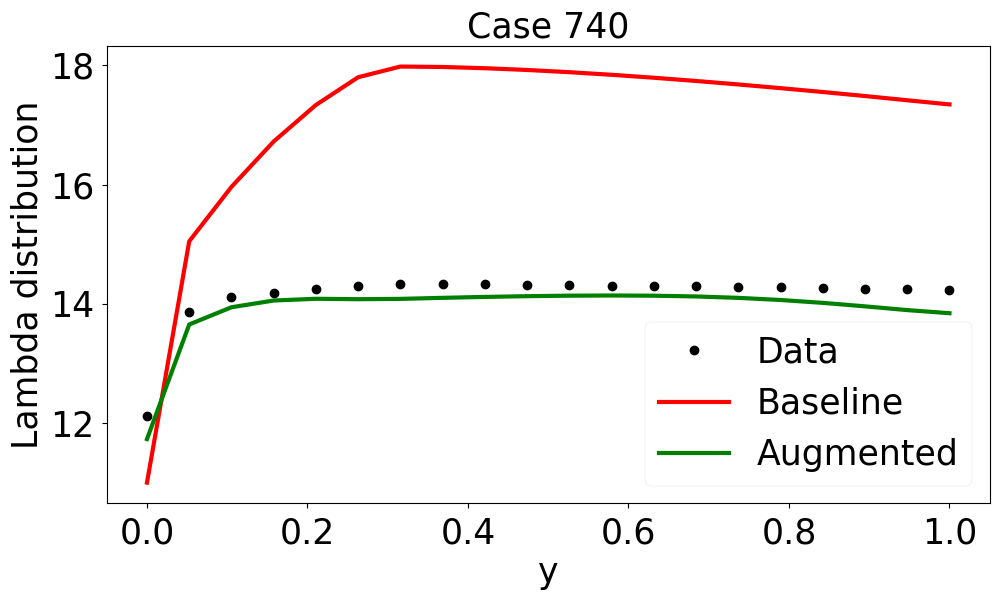}}
    \subfigure[$\mathscr{P}_1^\lambda=0.75$, $\mathscr{P}_2^\lambda=0.62$]
    {\includegraphics[width=0.31\textwidth]{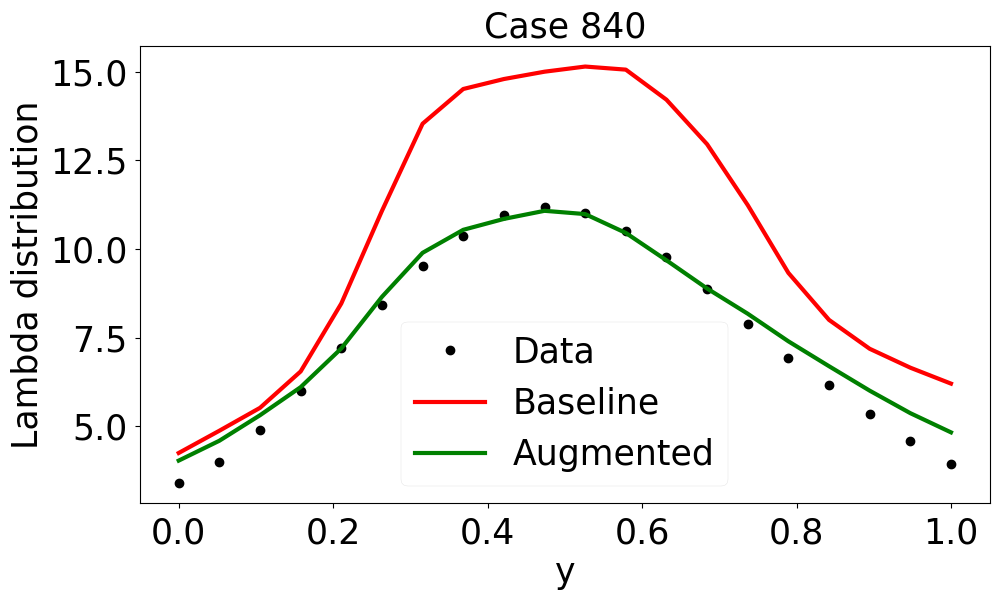}}
    \subfigure[$\mathscr{P}_1^\lambda=0.43$, $\mathscr{P}_2^\lambda=0.20$]
    {\includegraphics[width=0.31\textwidth]{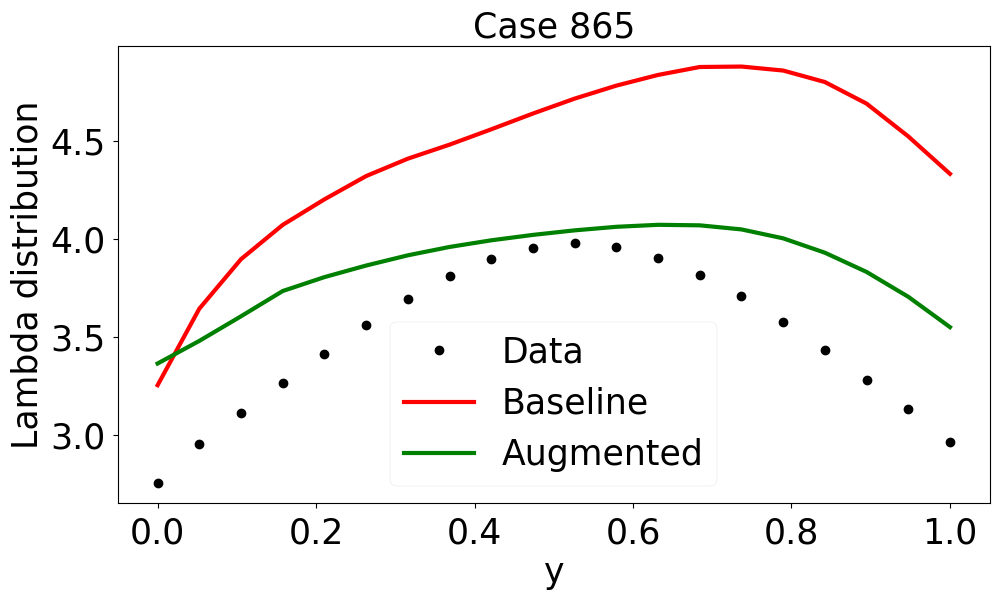}}
    \subfigure[$\mathscr{P}_1^\lambda=0.65$, $\mathscr{P}_2^\lambda=0.48$]
    {\includegraphics[width=0.31\textwidth]{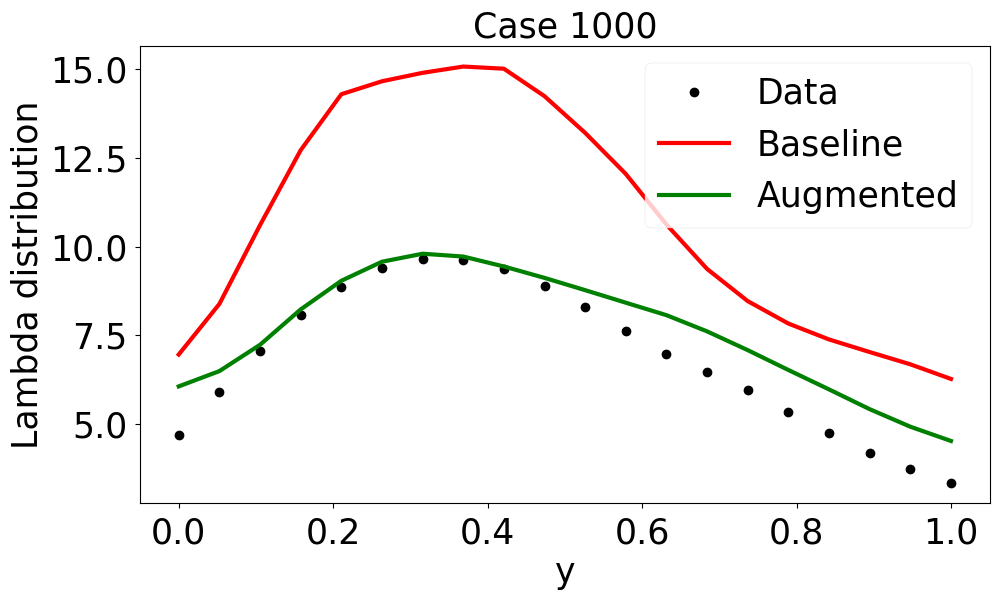}}
    \subfigure[$\mathscr{P}_1^\lambda=0.96$, $\mathscr{P}_2^\lambda=0.95$]
    {\includegraphics[width=0.31\textwidth]{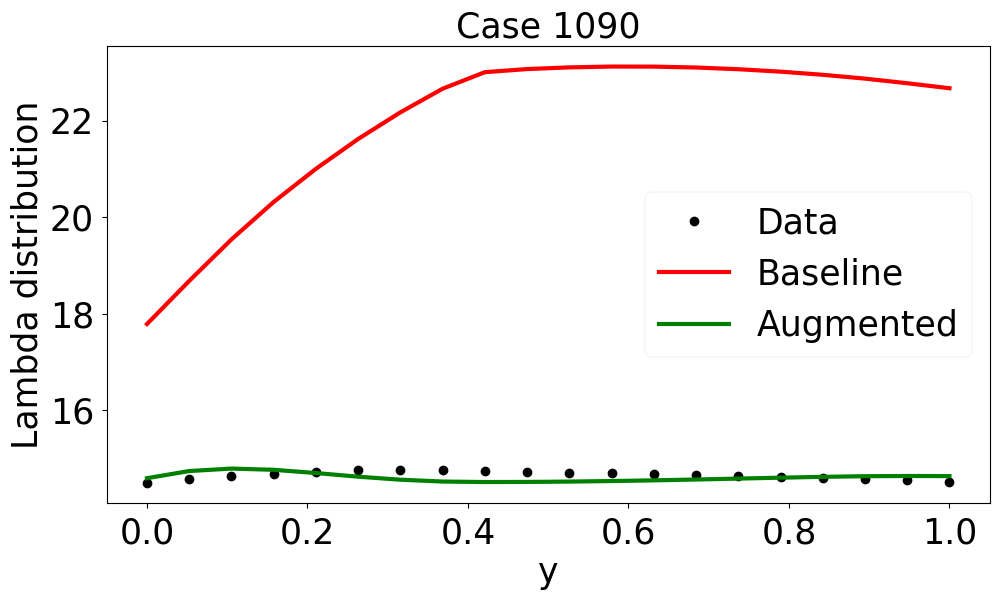}}
    \subfigure[$\mathscr{P}_1^\lambda=0.49$, $\mathscr{P}_2^\lambda=0.26$]
    {\includegraphics[width=0.31\textwidth]{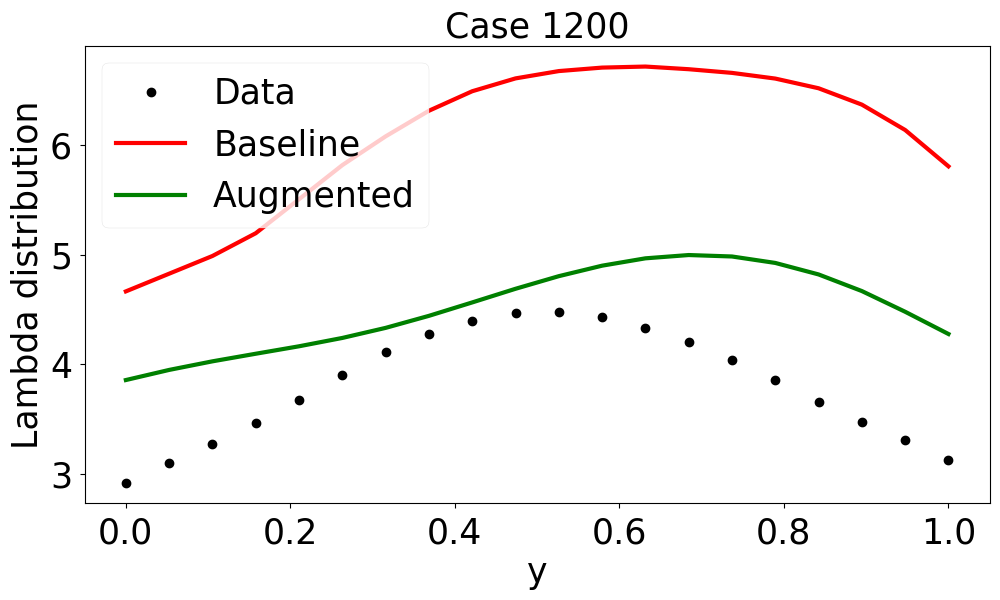}}
    \caption{Ionomer water content predictions for the training cases}
    \label{fig:training}
  \end{figure}
  
  The combined objective function for all 14 cases across inference iterations are shown in Fig. \ref{fig:optim}.
  The optimization could not proceed beyond iteration 23 because any subsequent augmentation function iterates caused the PyBaMM fuel cell model solver to diverge for some training cases (probably because of the increased numerical stiffness/instability introduced by the augmentation into the model).
  Predictive improvements in ionomer water content ($\lambda$) distributions w.r.t. available high-fidelity data for all training cases are plotted in Fig. \ref{fig:training}.
  As can be seen in the figure, some cases show very good improvements while some improved only marginally.
  A possible cause for this behavior could be the combined objective function being less sensitive to the feature-space regions where the features corresponding to the marginally improved cases lie.
  While a more careful choice of the training cases and the corresponding weights to the individual objective functions within the combined objective function might help, the objective here is to demonstrate the viability of the IIML approach to obtain generalizable improvements to the model.
  
  \subsection{Testing}
  
\begin{figure}[!h]
      \centering
      \includegraphics[width=1.0\textwidth]{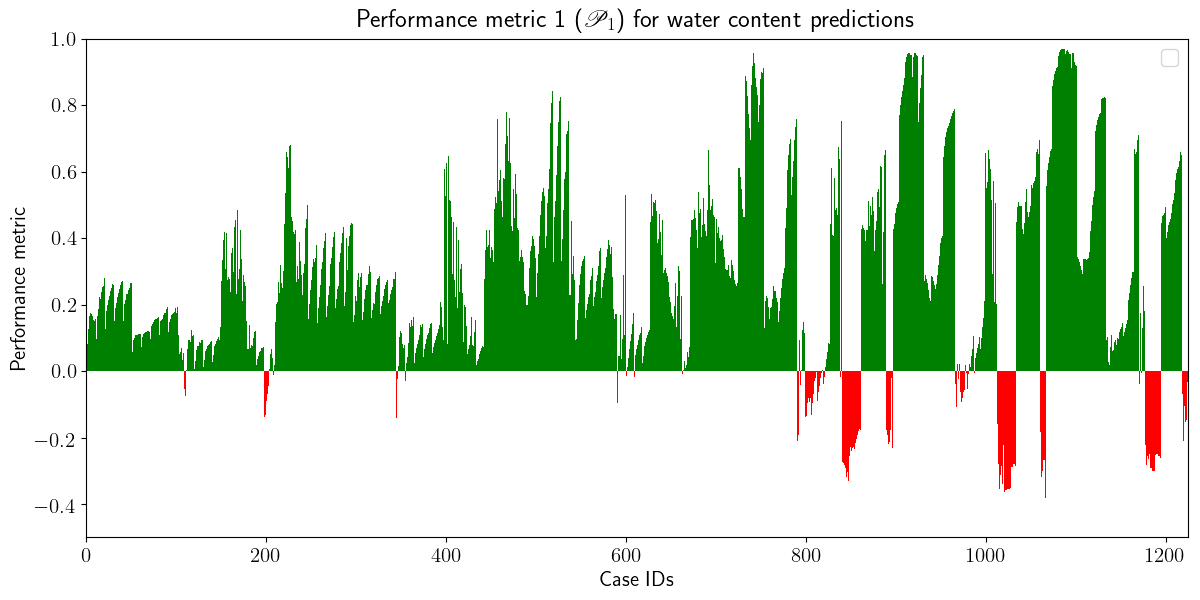}
      \caption{Performance metric $\mathscr{P}_1^\lambda$ for ionomer water content predictions across all 1224 cases}
      \label{fig:performance_water1}
  \end{figure}

\begin{figure}[!h]
      \centering
      \includegraphics[width=1.0\textwidth]{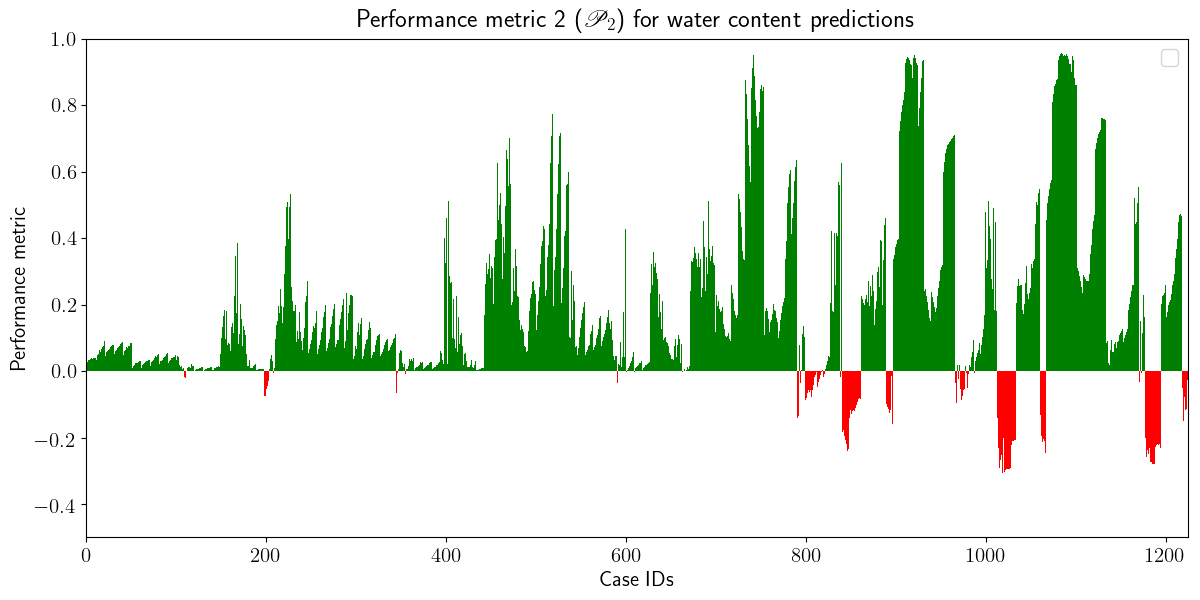}
      \caption{Performance metric $\mathscr{P}_2^\lambda$ for ionomer water content predictions across all 1224 cases}
      \label{fig:performance_water2}
  \end{figure}

  Once the training was completed, the resulting model was further tested over all available 1224 cases, the results for which are summarized in Figs. \ref{fig:performance_water1} and \ref{fig:performance_water2} using the following performance metrics, $\mathscr{P}_1$ and $\mathscr{P}_2$, which are defined for any quantity of interest $q$ as
  \begin{equation}
      \mathscr{P}_1^q = \frac{2\left\lVert q_\text{baseline} - q_\text{data} \right\rVert_2} {\left\lVert q_\text{augmented} - q_\text{data} \right\rVert_2 + \left\lVert q_\text{baseline} - q_\text{data} \right\rVert_2} - 1
  \end{equation}
  \begin{equation}
      \mathscr{P}_2^q = \frac{\mathscr{P}_1(q)\left\lVert q_\text{augmented} - q_\text{baseline} \right\rVert_2} {\left\lVert q_\text{augmented} - q_\text{data} \right\rVert_2 + \left\lVert q_\text{baseline} - q_\text{data} \right\rVert_2}.
  \end{equation}
  The performance metric $\mathscr{P}_1$, by design, is positive for cases where the augmented model gives a smaller $L_2$ error compared to the baseline model and negative when the error increases.
  The performance metric $\mathscr{P}_2$ scales $\mathscr{P}_1$ with the $L_2$ norm of the difference between the predictions from the augmented and baseline models.
  Thus, for a given case, $\lvert\mathscr{P}_2/\mathscr{P}_1\rvert\ll 1$ means that the baseline and augmented profiles are very close and that the baseline profile was reasonably accurate in the first place.
  On the other hand, if $\mathscr{P}_2/\lvert\mathscr{P}_1\rvert$ is close to unity, it means that the augmented model predicts accurately w.r.t. data and that the baseline model was significantly inaccurate compared to the augmented model.
  \begin{figure}[!h]
    \centering
    \subfigure[$\mathscr{P}_1^\lambda=0.27$, $\mathscr{P}_2^\lambda=0.09$]
    {\includegraphics[width=0.31\textwidth]{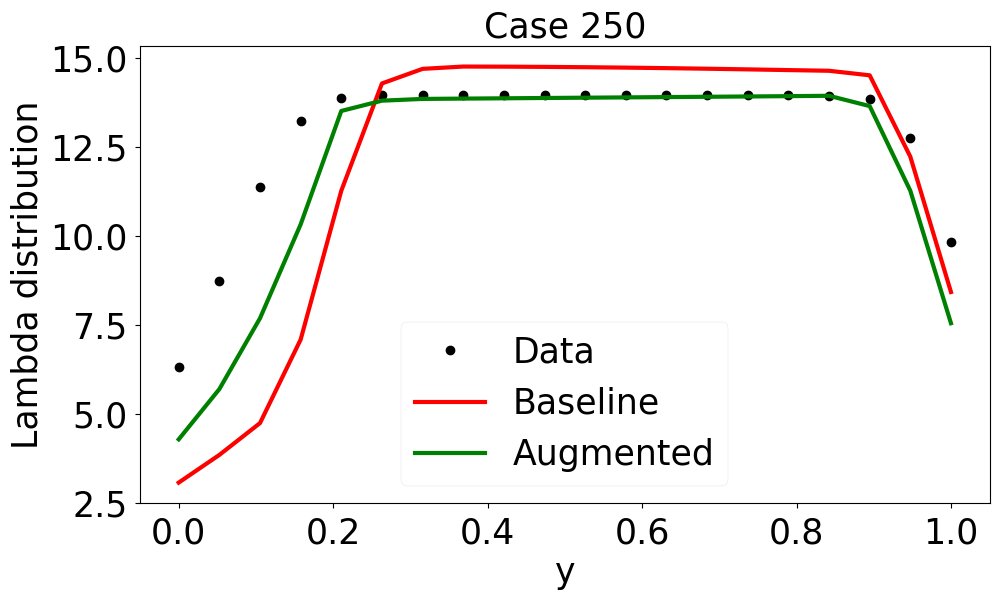}}
    \subfigure[$\mathscr{P}_1^\lambda=0.37$, $\mathscr{P}_2^\lambda=0.32$]
    {\includegraphics[width=0.31\textwidth]{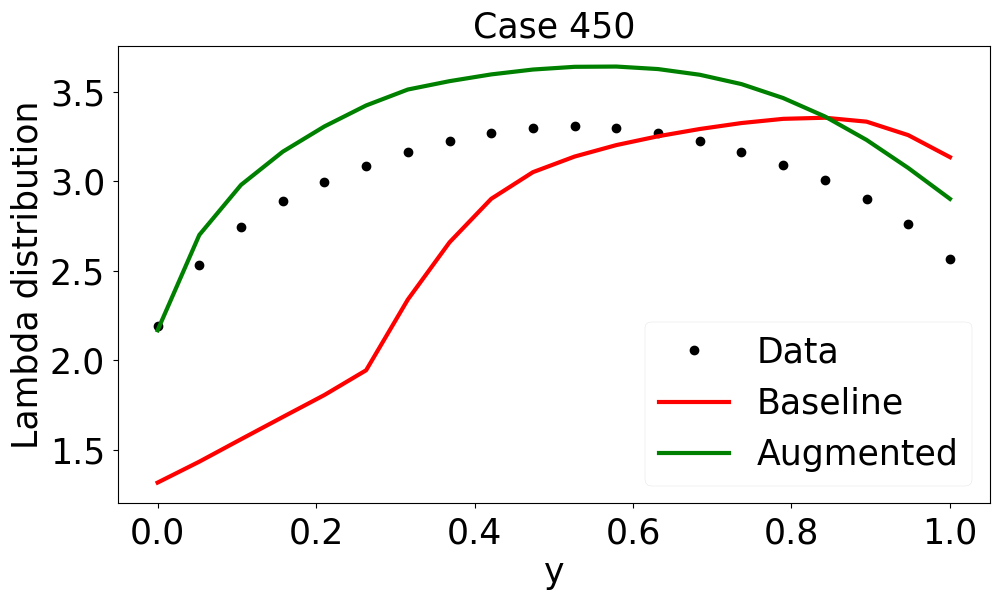}}
    \subfigure[$\mathscr{P}_1^\lambda=0.30$, $\mathscr{P}_2^\lambda=0.21$]
    {\includegraphics[width=0.31\textwidth]{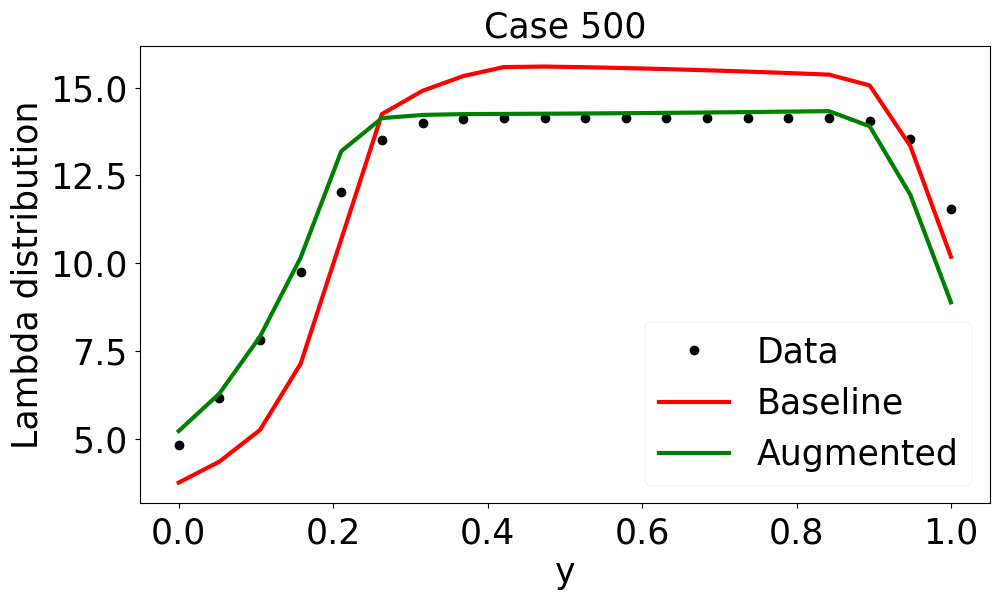}}
    \subfigure[$\mathscr{P}_1^\lambda=0.48$, $\mathscr{P}_2^\lambda=0.36$]
    {\includegraphics[width=0.31\textwidth]{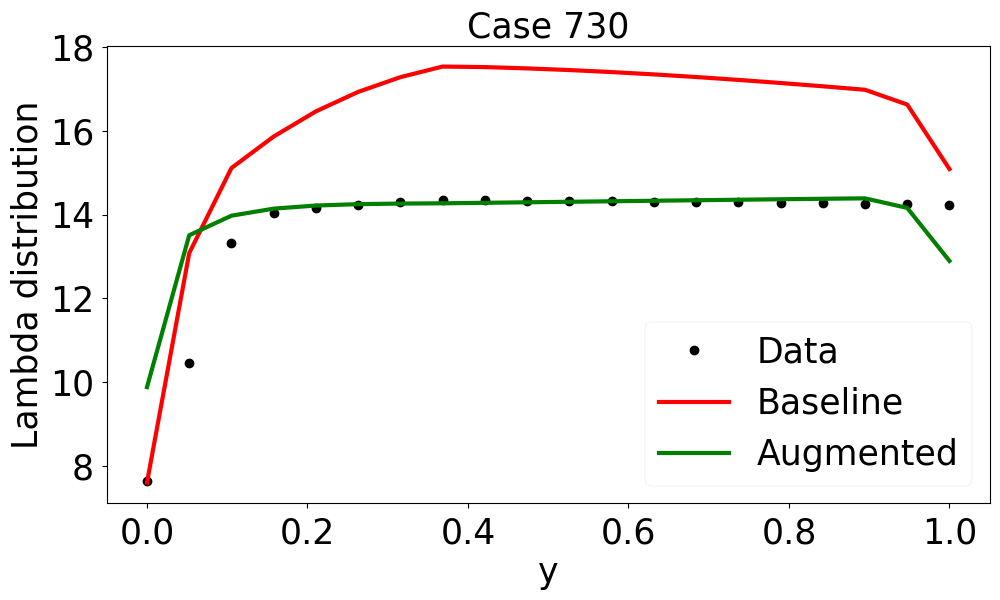}}
    \subfigure[$\mathscr{P}_1^\lambda=0.94$, $\mathscr{P}_2^\lambda=0.94$]
    {\includegraphics[width=0.31\textwidth]{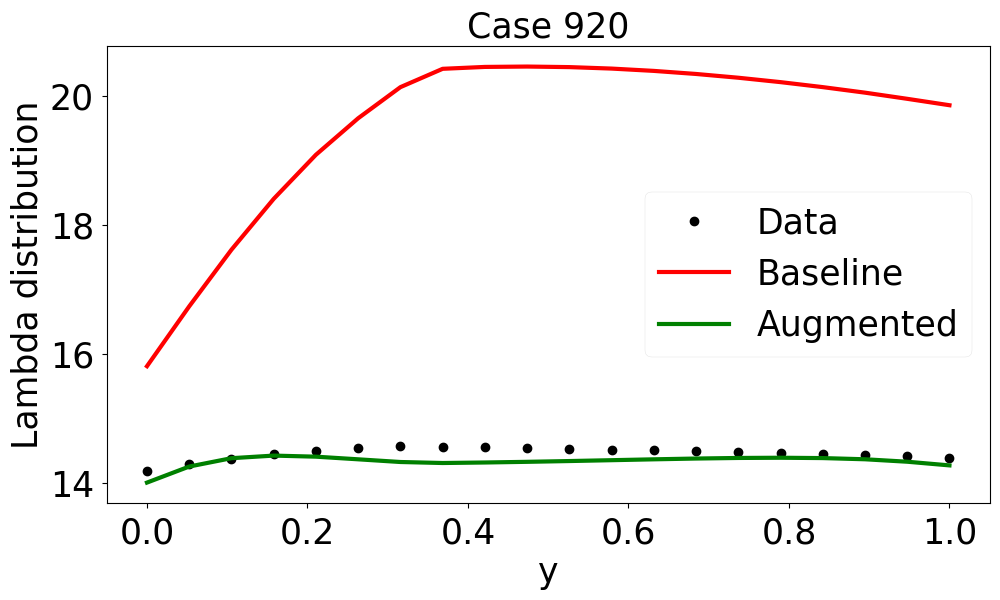}}
    \subfigure[$\mathscr{P}_1^\lambda=0.63$, $\mathscr{P}_2^\lambda=0.47$]
    {\includegraphics[width=0.31\textwidth]{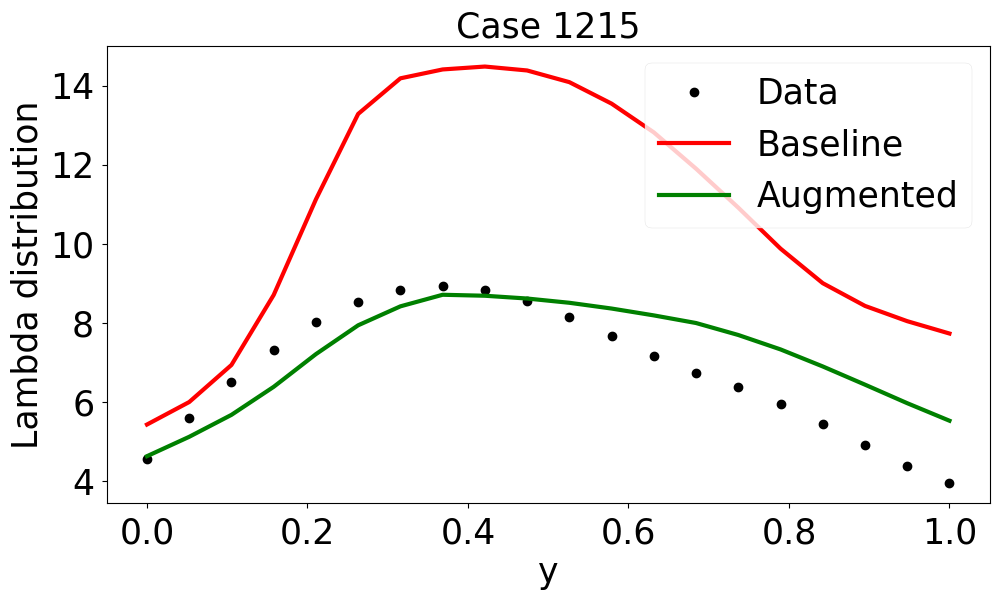}}
    \caption{Ionomer water content predictions for cases with high $\mathscr{P}_1^\lambda$ performance metrics}
    \label{fig:better_results}
  \end{figure}
  \begin{figure}[!h]
    \centering
    \subfigure[$\mathscr{P}_1^\lambda=-0.07$, $\mathscr{P}_2^\lambda=-0.02$]
    {\includegraphics[width=0.31\textwidth]{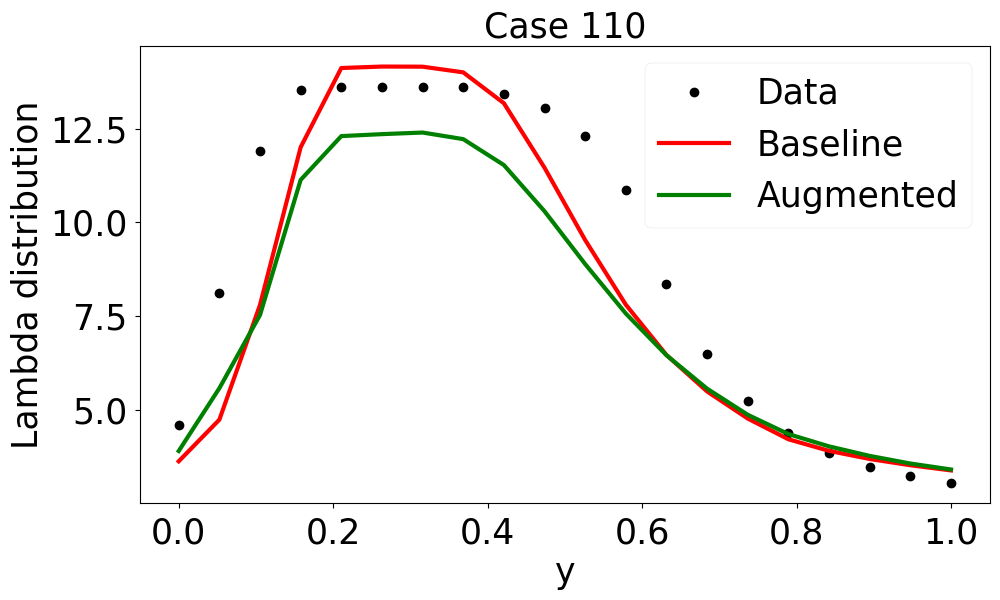}}
    \subfigure[$\mathscr{P}_1^\lambda=-0.05$, $\mathscr{P}_2^\lambda=-0.01$]
    {\includegraphics[width=0.31\textwidth]{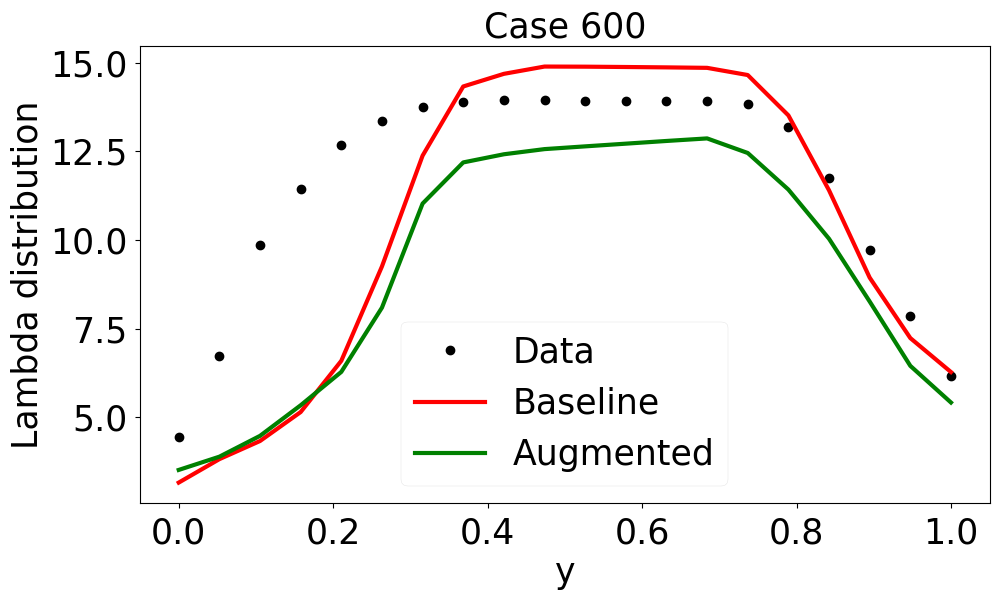}}
    \subfigure[$\mathscr{P}_1^\lambda=-0.02$, $\mathscr{P}_2^\lambda=-0.01$]
    {\includegraphics[width=0.31\textwidth]{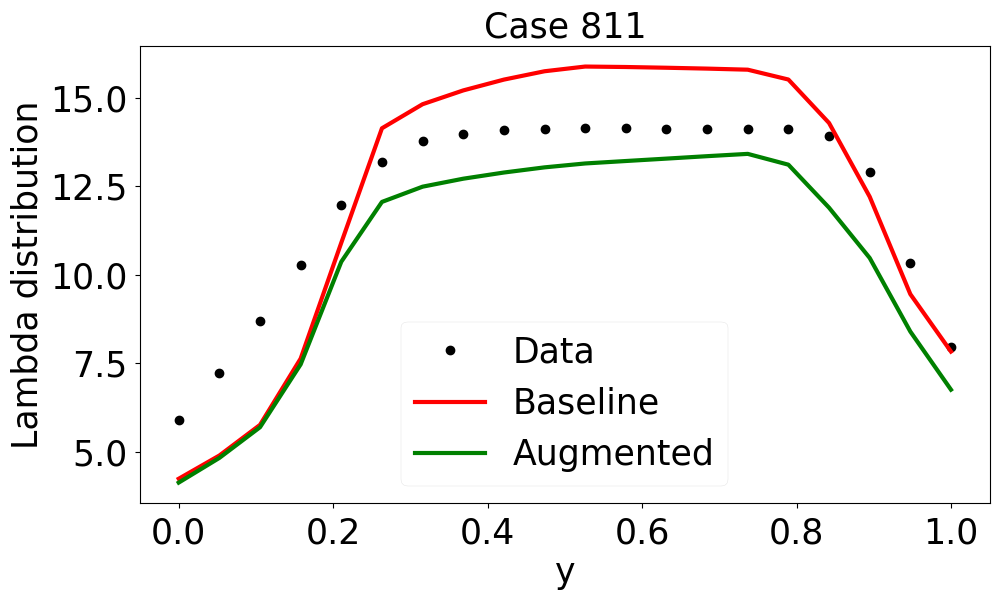}}
    \caption{Ionomer water content predictions for cases with $\mathscr{P}_1^\lambda$ values closest to zero}
    \label{fig:similar_results}
  \end{figure}
  \begin{figure}[!h]
    \centering
    \subfigure[$\mathscr{P}_1^\lambda=-0.33$, $\mathscr{P}_2^\lambda=-0.23$]
    {\includegraphics[width=0.31\textwidth]{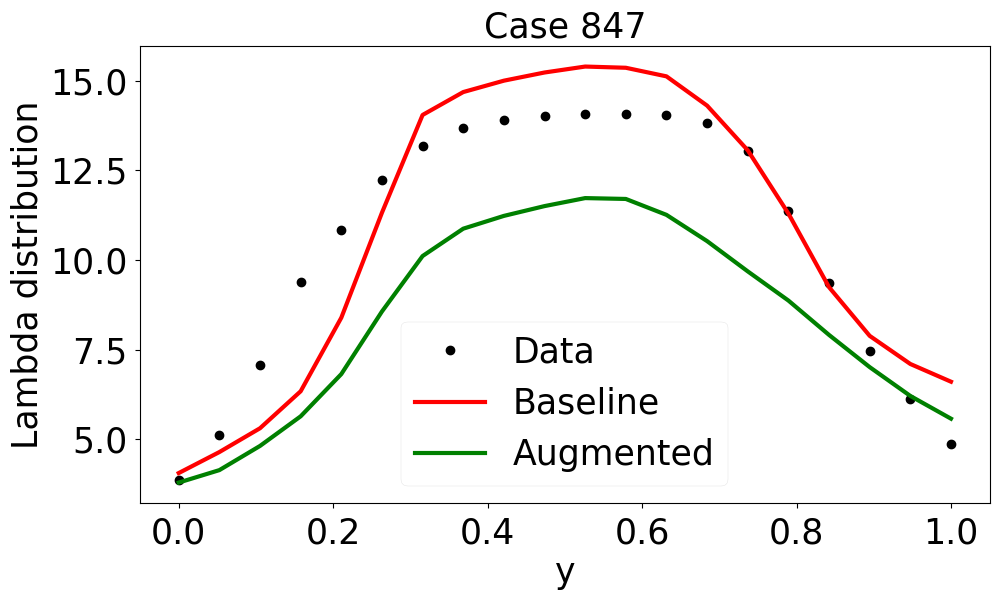}}
    \subfigure[$\mathscr{P}_1^\lambda=-0.36$, $\mathscr{P}_2^\lambda=-0.29$]
    {\includegraphics[width=0.31\textwidth]{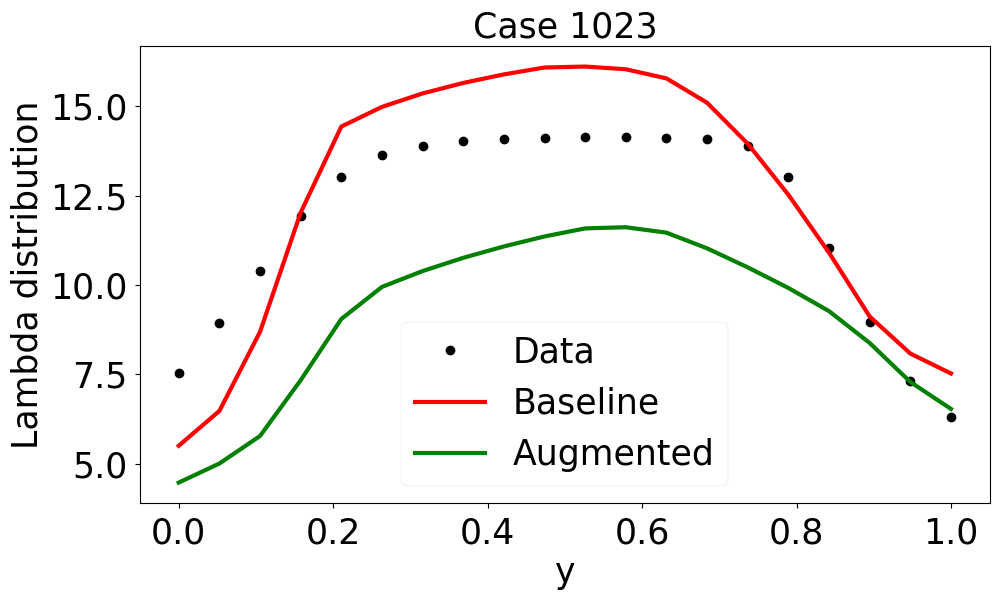}}
    \subfigure[$\mathscr{P}_1^\lambda=-0.38$, $\mathscr{P}_2^\lambda=-0.25$]
    {\includegraphics[width=0.31\textwidth]{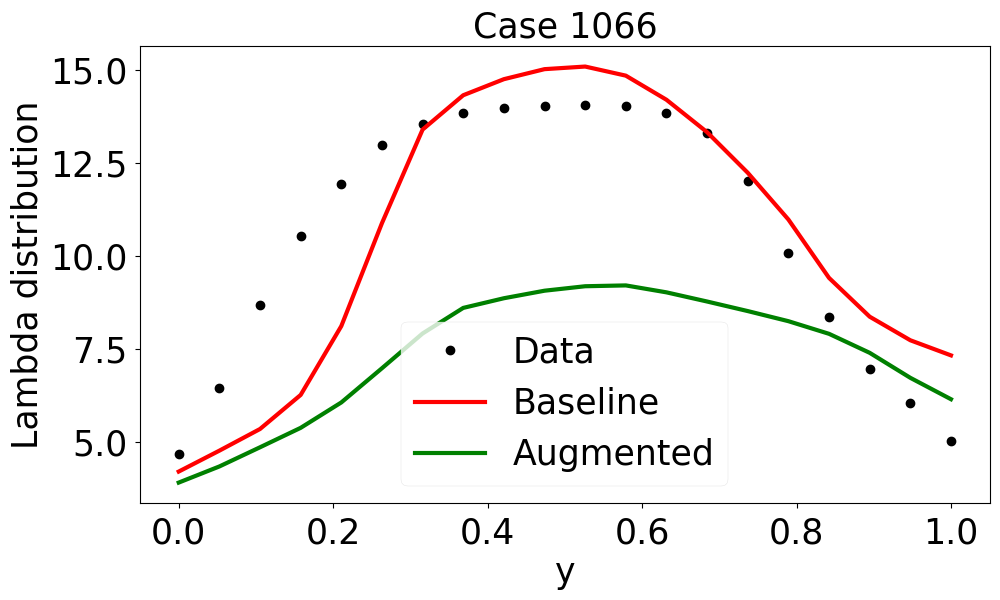}}
    \caption{Ionomer water content predictions for cases with low $\mathscr{P}_1^\lambda$ performance metrics. Conditions correspond to low cathode stoichiometry, high temperature, and low anode inlet relative humidity where the augmented model over-predicts drying of the cell (Low membrane water content)}
    \label{fig:worse_results}
  \end{figure}
  
  The cases where accuracy has improved are shown in green whereas the cases where it has deteriorated are shown in red.
  For 1087 out of 1224 cases the augmented model resulted in a lower L2 error compared to the baseline as indicated by a higher $\mathscr{P}_1$ metric.
  Figs. \ref{fig:better_results}, \ref{fig:similar_results} and \ref{fig:worse_results} show representative results associated with highly improved, marginally different, and significantly deteriorated $\mathscr{P}_1$ performance metrics.
  As can be seen in the results, the model seems to improve the predictions for a range of different physical conditions after training on just 14 representative cases.
  It should be noted that most cases with nearly zero or negative performance metrics exhibit a plateau in the spatial membrane water content distributions which indicates a saturated flow.
  Also, note that for some cases with values of $\mathscr{P}_1$ close to zero, the predictions are significantly different while predicting more accurately in one part of the physical space while falling short of even the baseline model in others.
  The plateau physically corresponds to saturated relative humidity conditions in the anode and cathode catalyst layers, (and possibly the channels) which drastically changes the water transport mechanism from a diffusion driven flux, to a mechanism of condensation and capillary flow.
  The relative difference in density of the gas vs liquid is $10^3$ which might explain why additional augmentations are needed to capture this highly nonlinear change in system behaviour.
  
\begin{figure}[!h]
      \centering
      \includegraphics[width=1.0\textwidth]{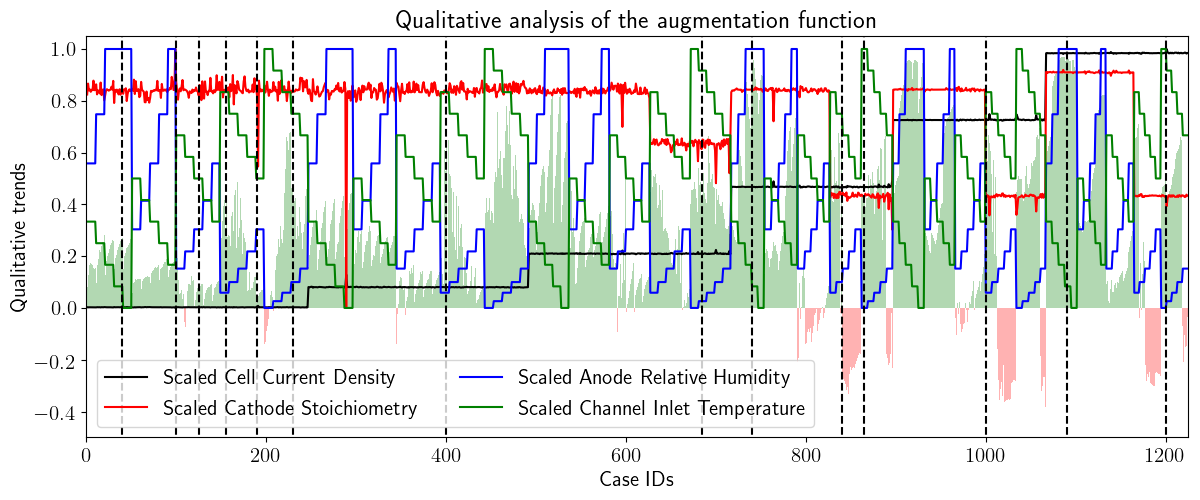}
      \caption{Qualitative trends of inflow conditions across different cases overlaid on $\mathscr{P}_1^\lambda$ trends (Black dashed lines indicate cases used for training).
      The performance of the model augmentation is highly dependent on both the anode inlet RH, Temperature and operating current, and the worst performance corresponds to high current, low temperature and low anode inlet RH conditions.}
      \label{fig:analysis1}
  \end{figure}

\begin{figure}[!h]
      \centering
      \includegraphics[width=1.0\textwidth]{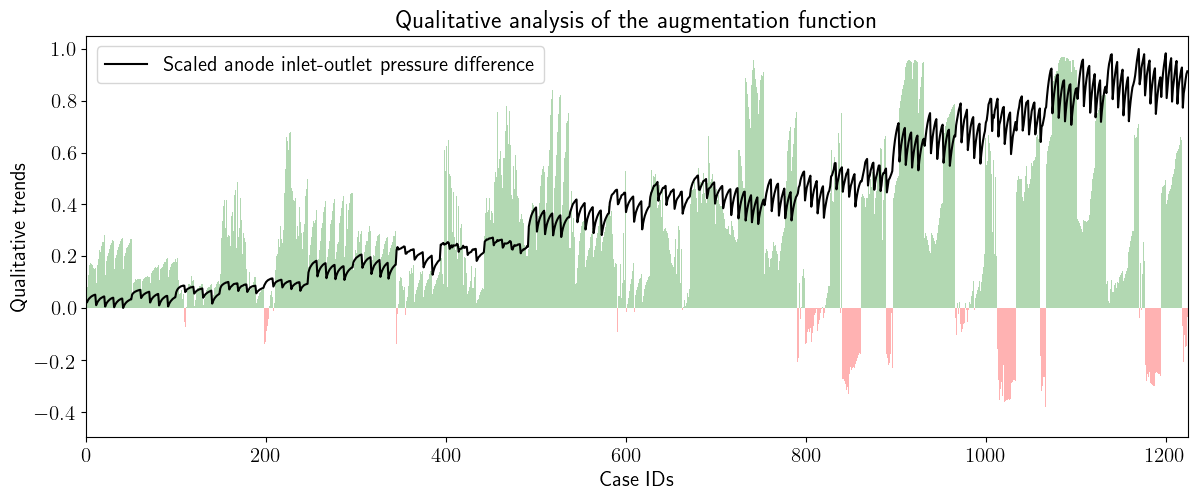}
      \caption{Qualitative trends of anode pressure difference across different cases overlaid on $\mathscr{P}_1^\lambda$ trends}
      \label{fig:analysis2}
  \end{figure}

  Figs. \ref{fig:analysis1} and \ref{fig:analysis2} show the qualitative trends of inflow and outflow conditions overlaid on the evaluated performance metrics.
  Empirical evidence suggests that the augmented model performs significantly worse than its baseline counterpart under two specific sets of inflow/outflow conditions, both of which are characterized by high cell current densities.
  The first set of conditions involves a high cathode stoichiometric ratio, high anode relative humidity and low temperature and the second set of conditions involves a low cathode stoichiometric ratio, low relative humidity and high temperature.
  Here, stoichiometric ratio refers to the ratio of inlet oxygen to reacted oxygen.
  High cathode stoichiometric ratio is important because these cases increase the water removal rate from the cathode inlet area causing larger non-uniformity and partially saturated conditions along the channel due to higher water generation rate with higher currents.
  Additionally, the sharp changes in performance metrics across consecutive case numbers is consistent with the variation of pressure difference between anode inlet and outlet.
  This can be seen from the high frequency oscillations with respect to case numbers in Fig. \ref{fig:analysis2}.
  
  Given the complex interactions between various sub-models within the fuel-cell model itself and such a high-dimensional feature-space, a model would require a highly intricate functional form and a large amount of data to make accurate predictions for any arbitrary inflow conditions, if such predictions are possible at all.
  
  
  \subsection{Changes in Current Density Predictions}
  
  \begin{figure}[!h]
      \centering
      \includegraphics[width=1.0\textwidth]{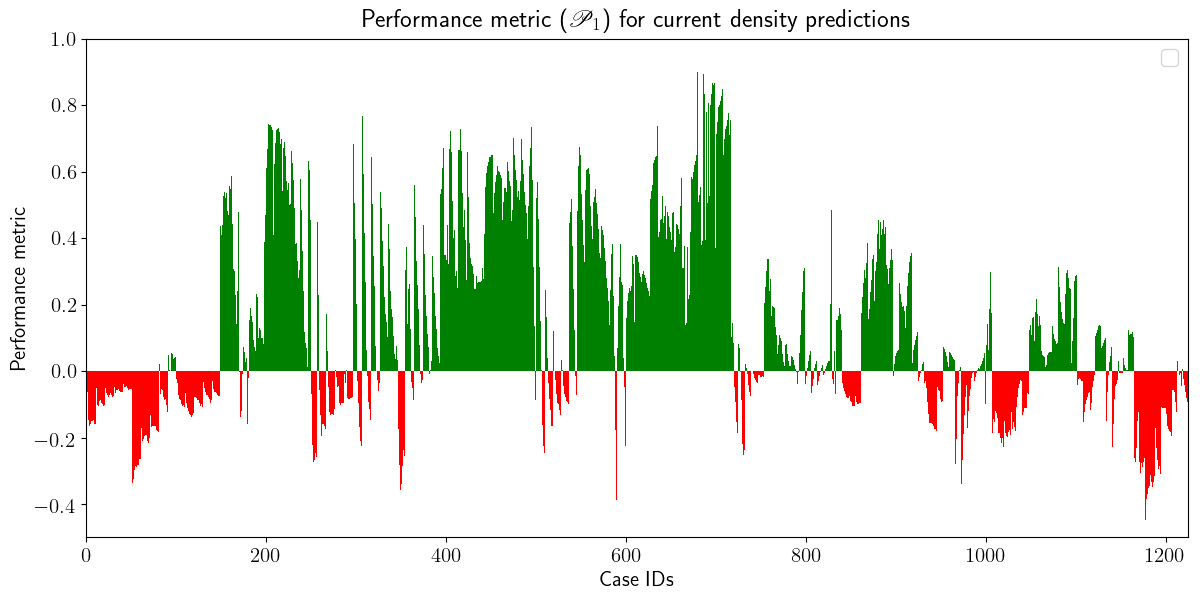}
      \caption{Performance metric $\mathscr{P}_1^j$ for current density predictions across all 1224 cases}
      \label{fig:performance_current1}
  \end{figure}
  \begin{figure}[!h]
      \centering
      \includegraphics[width=1.0\textwidth]{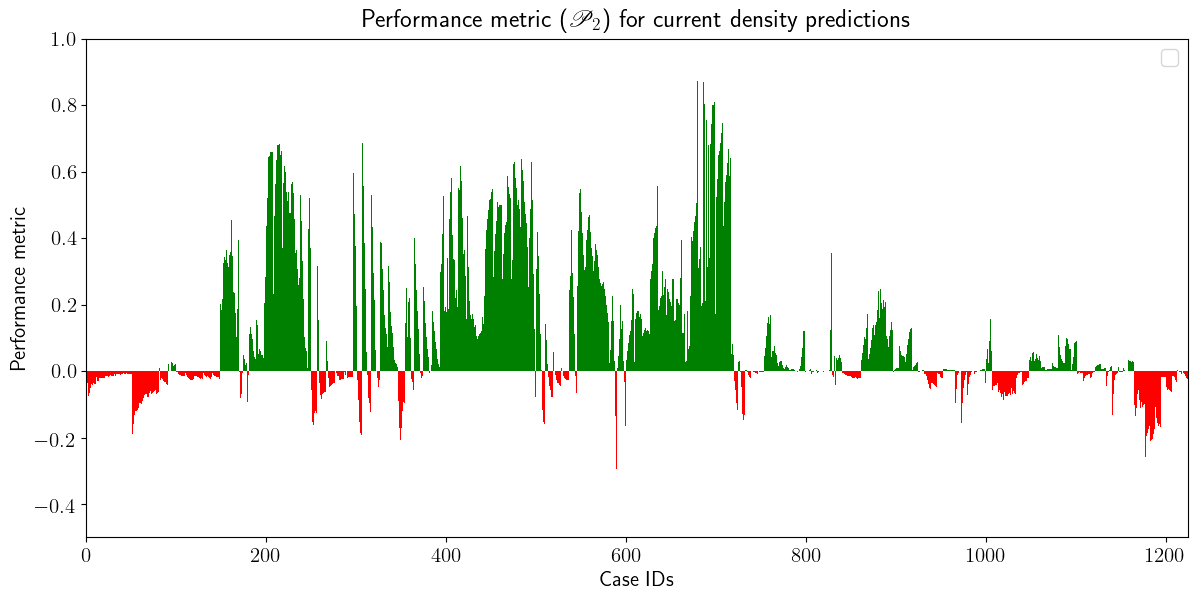}
      \caption{Performance metric $\mathscr{P}_2^j$ for current density predictions across all 1224 cases}
      \label{fig:performance_current2}
  \end{figure}
  
  \begin{figure}[!h]
    \centering
    \subfigure[$\mathscr{P}_1^j=0.74$, $\mathscr{P}_2^j=0.64$]
    {\includegraphics[width=0.31\textwidth]{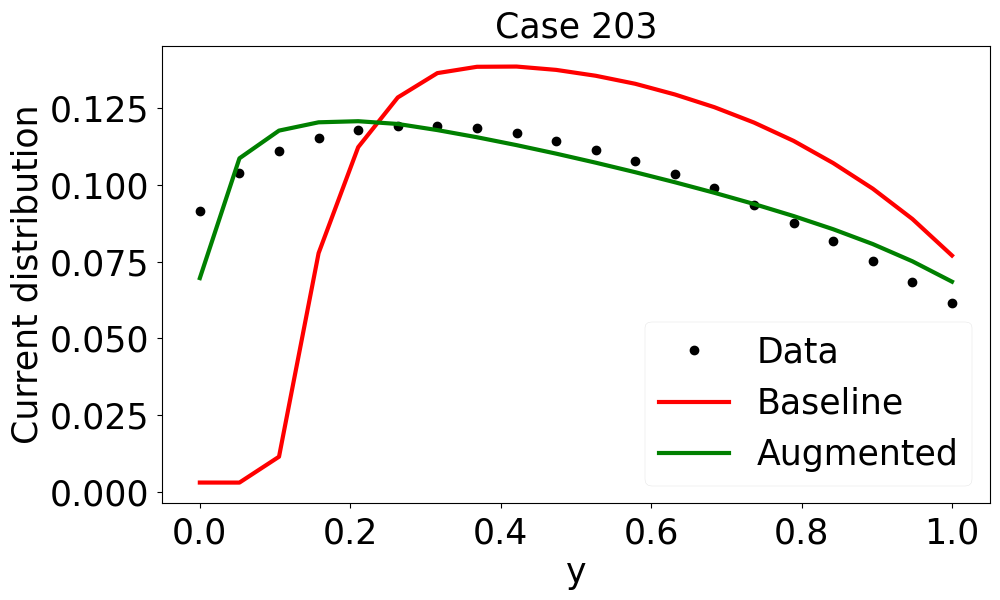}}
    \subfigure[$\mathscr{P}_1^j=0.65$, $\mathscr{P}_2^j=0.55$]
    {\includegraphics[width=0.31\textwidth]{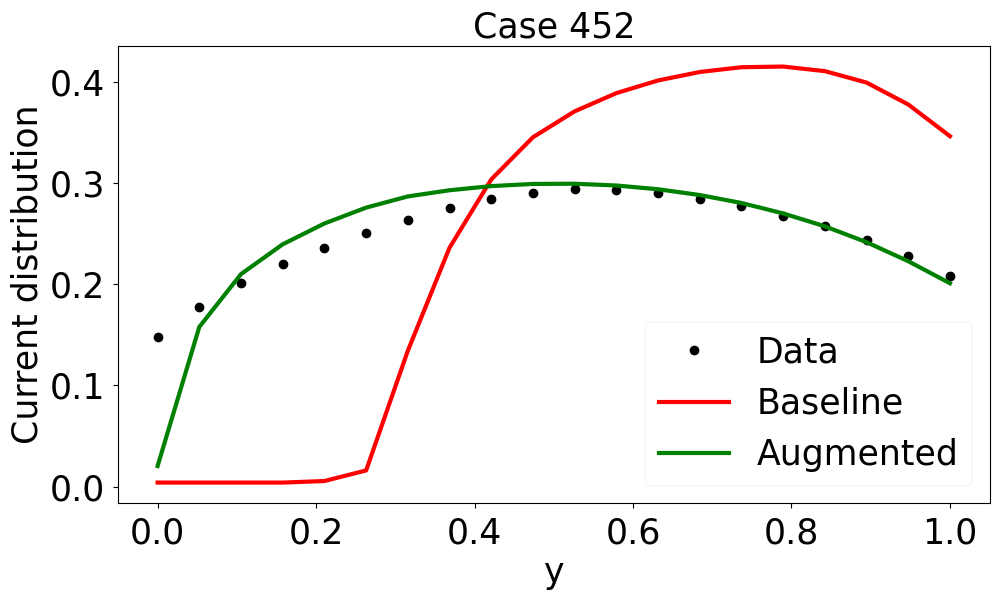}}
    \subfigure[$\mathscr{P}_1^j=0.53$, $\mathscr{P}_2^j=0.34$]
    {\includegraphics[width=0.31\textwidth]{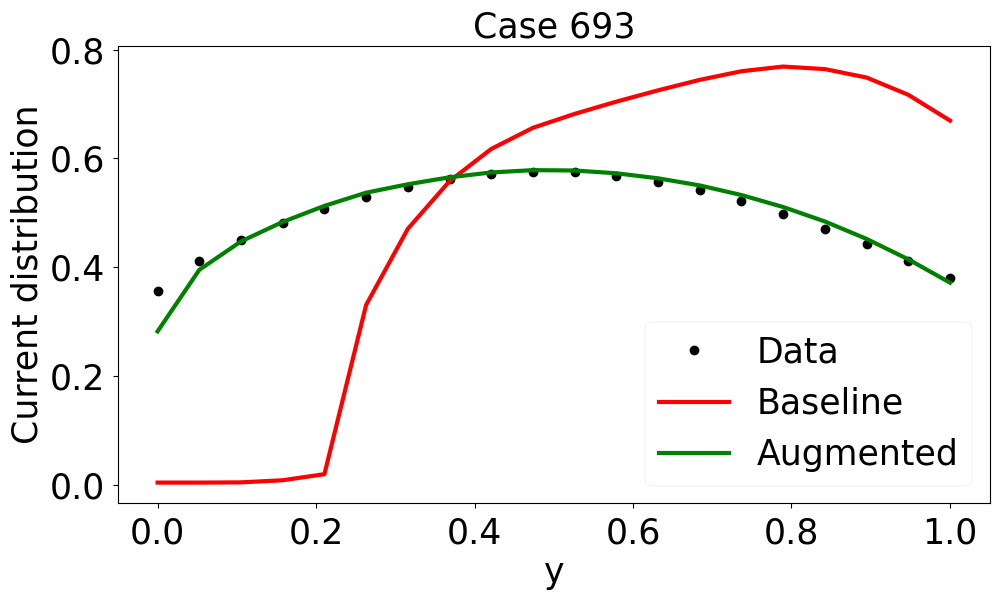}}
    \subfigure[$\mathscr{P}_1^\lambda=-0.04$, $\mathscr{P}_2^\lambda=-0.03$]
    {\includegraphics[width=0.31\textwidth]{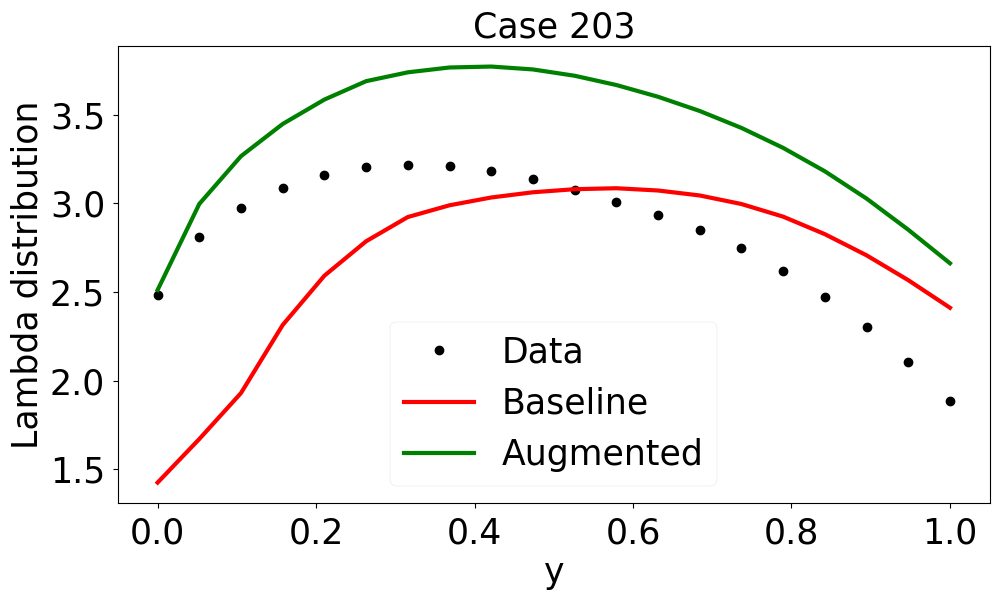}}
    \subfigure[$\mathscr{P}_1^\lambda=0.36$, $\mathscr{P}_2^\lambda=0.31$]
    {\includegraphics[width=0.31\textwidth]{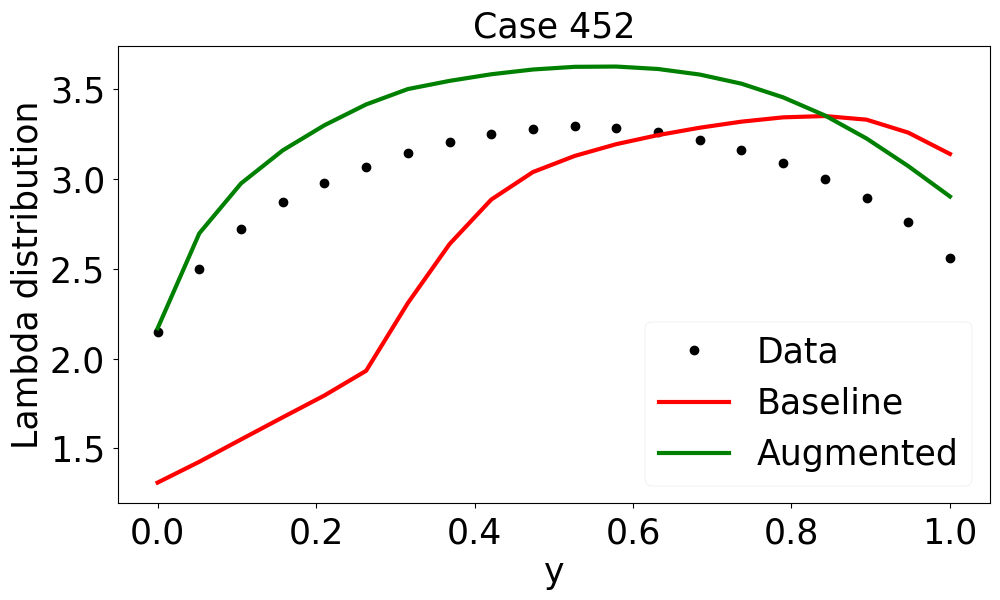}}
    \subfigure[$\mathscr{P}_1^\lambda=0.56$, $\mathscr{P}_2^\lambda=0.37$]
    {\includegraphics[width=0.31\textwidth]{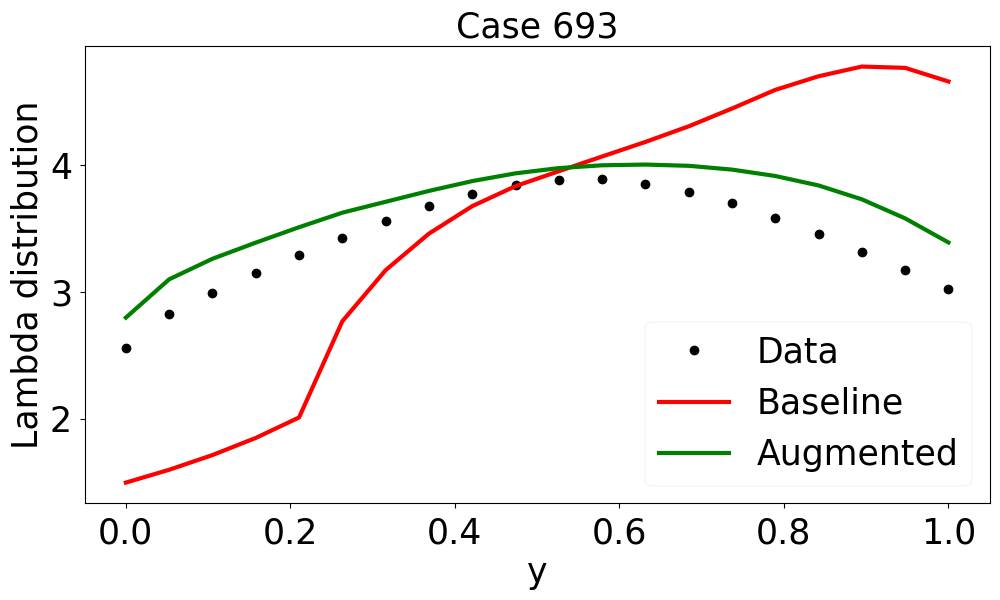}}
    \caption{Current density and water content predictions for cases with high $\mathscr{P}_1^j$ performance metrics}
    \label{fig:better_results_current}
  \end{figure}
  \begin{figure}[!h]
    \centering
    \subfigure[$\mathscr{P}_1^j=-0.34$, $\mathscr{P}_2^j=-0.18$]
    {\includegraphics[width=0.31\textwidth]{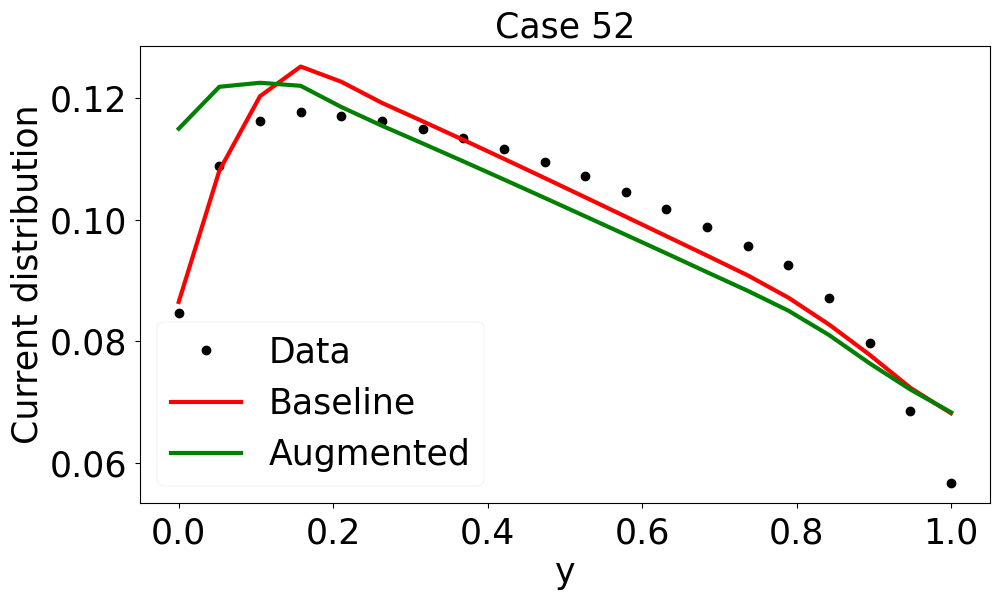}}
    \subfigure[$\mathscr{P}_1^j=-0.17$, $\mathscr{P}_2^j=0.07$] 
    {\includegraphics[width=0.31\textwidth]{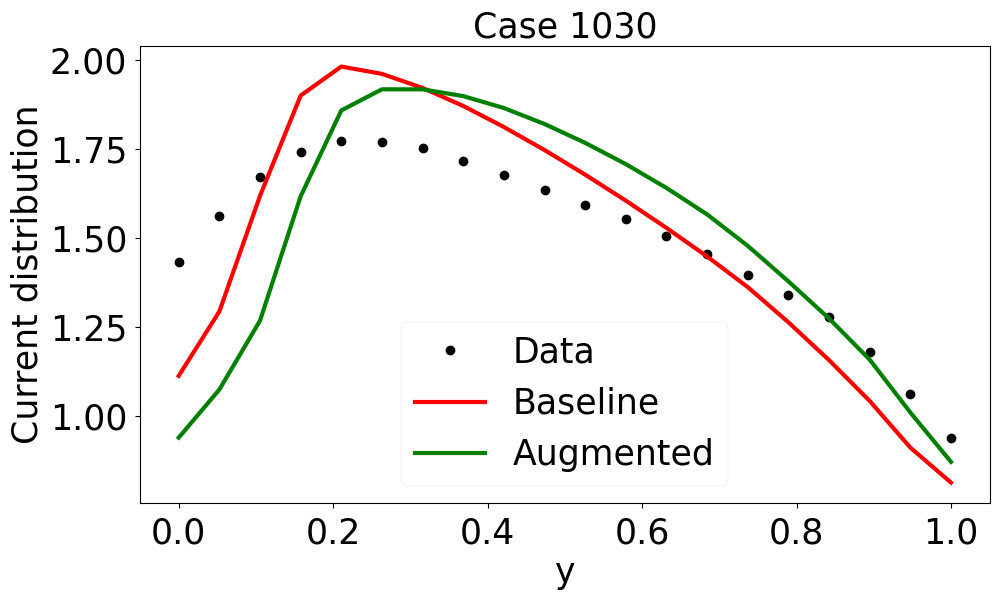}}
    \subfigure[$\mathscr{P}_1^j=-0.11$, $\mathscr{P}_2^j=-0.02$]
    {\includegraphics[width=0.31\textwidth]{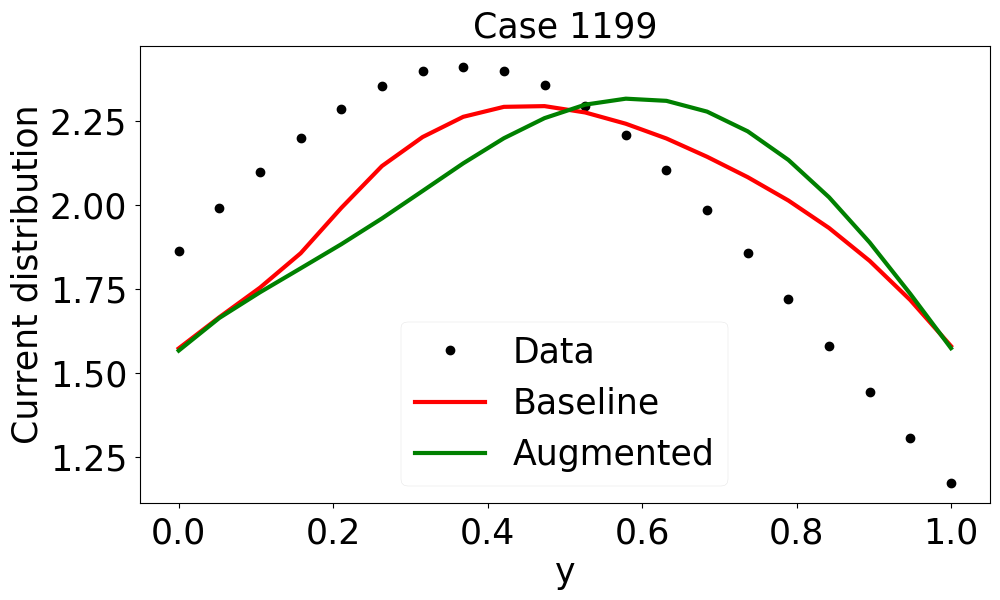}}
    \subfigure[$\mathscr{P}_1^\lambda=0.08$, $\mathscr{P}_2^\lambda=0.01$]
    {\includegraphics[width=0.31\textwidth]{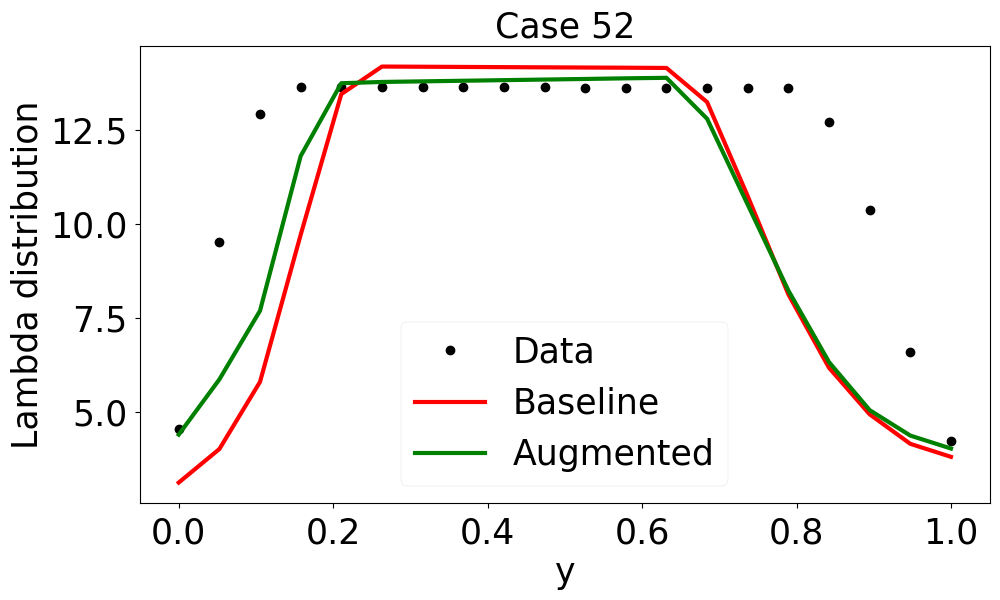}}
    \subfigure[$\mathscr{P}_1^\lambda=-0.29$, $\mathscr{P}_2^\lambda=-0.21$] 
    {\includegraphics[width=0.31\textwidth]{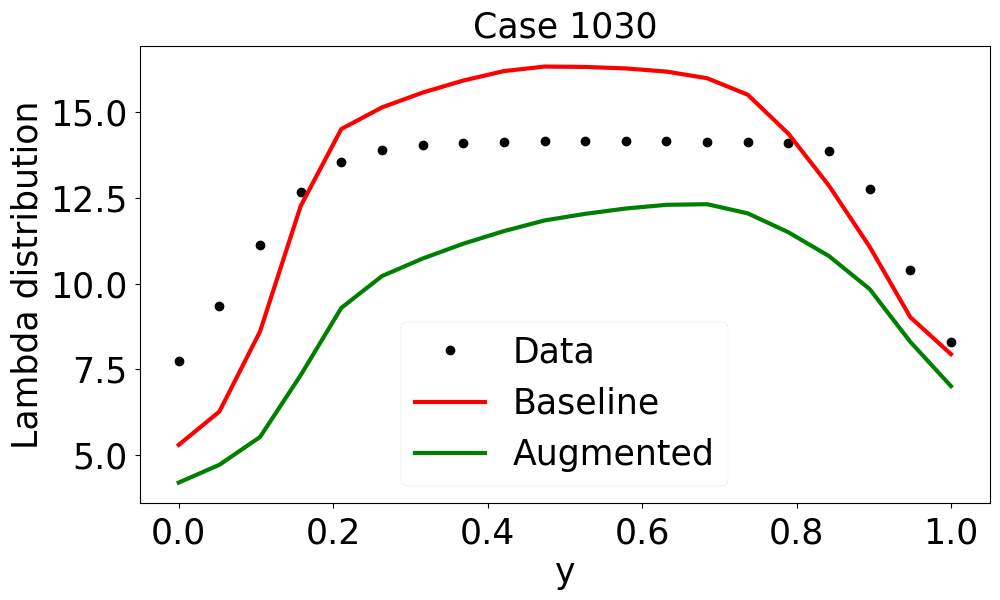}}
    \subfigure[$\mathscr{P}_1^\lambda=0.48$, $\mathscr{P}_2^\lambda=0.23$]
    {\includegraphics[width=0.31\textwidth]{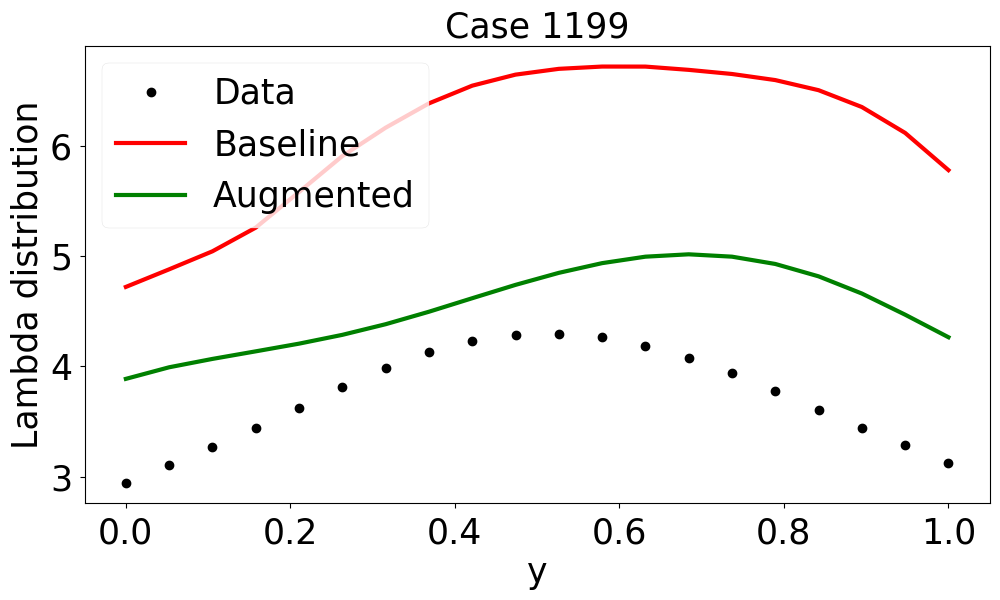}}
    \caption{Current density and water content predictions for cases with low $\mathscr{P}_1^j$ performance metrics}
    \label{fig:worse_results_current}
  \end{figure}
  To judge the quality of predictions for other physical quantities, individual comparisons for the current density distributions are presented for a few selected cases which show better and worse results compared to the corresponding high-fidelity data are shown in Figs. \ref{fig:better_results_current} and \ref{fig:worse_results_current}.
  The performance metrics w.r.t. the predictions for current density distributions are summarized in Figs. \ref{fig:performance_current1} and \ref{fig:performance_current2} across all of the 1224 cases.
  Since the current density is not the intended output of the augmented model the presented results are not completely unexpected.
  However, it should be noted here that even for several cases with fairly low $\mathscr{P}_1$, $\mathscr{P}_2$ is significantly smaller in magnitude, i.e. the predictions from the augmented model are close to those from the baseline model.
  Thus, for most cases, the augmented model either stays close to the baseline model or improves it.
  For several cases with high performance metrics, we do see a significant correction in the current density predictions.
  This result is expected because the membrane water content impacts the proton conductivity as shown in Eqn. \ref{eqn:proton_charge_balance}, and hence the current density distribution should follow the shape of the membrane water content unless the stoichiometric ratio is very low, or the temperature gradient along the y direction is high.
  Finally, note that for a few cases (e.g. case 1199), even though the prediction error for the ionomer water content decreases, even the qualitative trends for the water content are wrong, and correspondingly, the current density predictions also contain significant errors when compared to the high-fidelity data.
  Further work and analysis is needed to ascertain whether such cases require a different treatment during the inference process, or a different physical augmentation point in the model.
  
  The impact of using less data than in the above experiments can be found in Appendix B.

%% file: 5_Conclusions/main.tex
\section{Conclusions} \label{sec:Conclusions}

  The weakly-coupled Integrated Inference and Machine Learning (IIML) framework is presented which enables inference of data-driven model-consistent augmentations.
  Weakly-coupled IIML constrains the inference problem to a learnable manifold and enforces consistency among features-to-augmentation maps across the training dataset.
  This is achieved by performing a machine learning step after every inference update made to the spatial field of augmentation values.
  To maintain consistency with the learned augmentation, a field correction step, consisting of a forward solve using the augmented model, is carried out before starting the next inference update.
  When used with an iterative solution strategy, this framework removes the requirement to embed the augmentation function within the numerical solver.
  The only changes that need to be made are the addition of a spatial field of augmentation values as an array and applying the appropriate augmentation values within the numerics as required.
  This can significantly reduce the time and effort required to setup an inference problem at the expense of increased computation time (owing to iterative solution of the augmented model) to solve the problem.
  However, this trade-off is acceptable when working with reduced order (fast-running) models.

  The weakly-coupled IIML framework was used to augment an existing linearized 1+1D proton exchange membrane fuel cell (PEMFC) model in order to better predict the x-averaged membrane water content distribution along the channel length (y-direction).
  To introduce the augmentation term into the model, the equilibrium membrane water content ($\lambda_\text{eq}$) function was multiplied by an augmentation function $\beta$.
  The augmentation was assumed to be a function of eight features (which in turn are functions of model states) that include mole fractions of water vapor in the anode and cathode channels, water vapor concentrations and water content in the anode and cathode catalyst layers, membrane water content and temperature inside the cathode channel.
  
  The high-fidelity data for all 1224 different cases (operating conditions including flow rates, relative humidity, and power level) in the dataset was obtained from a proprietary 2D model of a fuel cell from the 2016 Toyota Mirai.
  To demonstrate generalizability, only 14 cases out of the available 1224 cases were chosen for training.
  The choice of the training cases was based on the need to expose the neural network to different types of physical phenomena that can take place due to significantly different boundary conditions.
  Since the objective here is to demonstrate the range of applicability of such augmentations, the number of training cases in this work was intentionally kept as low as possible to minimize similarity with most testing cases.
  
  Once trained, the predictive capability of the augmented model was demonstrated for the intended output, i.e. the membrane water content distribution and another output not involved in the inference process, viz. current density distribution.
  Performance metrics were designed to judge the capability of the augmented model relative to the baseline model.
  The membrane water content predictions improved for a majority (1087/1224) of cases which provides a testament to the capability of the framework to produce generalizable augmentations.
  However, a small fraction of cases showed predictions which were worse compared to the baseline model, and did not predict the respective water content distributions accurately.
  A more rigorous tuning of the neural network hyper-parameters, a more careful choice of the training cases, and a better design of the combined objective function might help in improving predictive capabilities of the current model.
  
  The changes in the current density predictions were more nuanced in the sense that even cases with a fairly low performance metric showed comparable predictions to the baseline model and large discrepancies were observed only in a small part of the domain (usually near the cathode air inlet y=0).
  However, for cases exhibiting high performance metrics, varying degrees of improvements were observed in the predictive accuracy.
  Hence, to some extent, the IIML procedure not only improved the ionomer water content predictions but the current density predictions as well.
  
  For comparative purposes, a second augmentation was also trained using only 7 out of the 14 configurations in the original training dataset.
  The results confirm that having too little data reduces the predictive accuracy of the augmented model across the testing cases.
  Although, using fewer training cases allows the inference process to overfit augmentation behavior specific to the training cases (and hence predict more accurately on the training cases), the resulting augmentation function generally yields less accurate predictions  when compared to a training dataset with more data.
  
  Overall, the application of the weakly-coupled IIML approach resulted in improved predictive accuracy of the fuel cell model.
  Further gains in the accuracy of an augmented model can be achieved by introducing more refined physical parametrizations.
  Along these lines, we remark that the main contribution of this study - the IIML approach - presents the modeler a new set of tools.

\section*{Acknowledgements}
  
  The authors acknowledge funding from Toyota Motor Engineering and Manufacturing North America. We thank Ken Butts and Oana Nitulescu for their insightful discussions.

%% file: A_SourcesAndBoundaryConditions/main.tex
\section{Fuel Cell Model Source Terms} \label{app:SourceTerms}

  The definitions of all source terms and boundary conditions for all equations used to model the fuel cell (see Section \ref{ssec:FuelCellModel}) have been given as follows.
  The Butler-Volmer relation is used to model the reaction-current density $j_{cl}$ induced by the half-reactions in the catalyst layers.
  \begin{equation}
    j_{cl} = i_0(c_k, T) \left(\exp\left(\frac{2\beta F}{RT}\eta\right)-\exp\left(-\frac{2(1-\beta)F}{RT}\eta\right)\right) \quad \text{where} \quad k\in\lbrace O_2,H_2\rbrace
  \end{equation}
  Here, $\eta$ is the overpotential given by
  \begin{equation}
    \eta = \phi_e - \phi_p - U(c_k,T) \quad \text{where} \quad k\in\lbrace O_2,H_2\rbrace
  \end{equation}
  $i_0$ is the exchange-current density and $U$ is the reversible potential difference, both of which are functions of temperature $T$ and the appropriate concentration ($c_{\text{H}_\text{2}}$ or $c_{\text{O}_\text{2}}$).
  $F$ is the Faraday constant.
  The sign convention used here assumes that $j_{cl}$ is positive at the anode (where the oxidation of hydrogen occurs).
  Since no reactions occur outside the catalyst layers, the interfacial current density can be written as follows.
  \begin{equation}
    j = \left\lbrace \begin{matrix} j_{cl}, & x\in\Omega_{cl}\\  0, & \text{otherwise}\end{matrix} \right.
  \end{equation}
  The rate of consumption of hydrogen and oxygen can be written in terms of the interfacial current density as follows.
  \begin{equation}
    r_{H_2} = -\frac{aj}{2F}, \qquad r_{O_2} = \frac{aj}{4F}
  \end{equation}
  The evaporation/condensation source term can be given as follows in terms of the water vapor concentration and saturation concentration (which is a function of the saturation pressure $p_\text{sat}$, which in turn varies with temperature).
  \begin{equation}
    S_{ec} = \gamma_{ec} (c_{\text{H}_\text{2}\text{O}} - c_\text{sat}), \qquad c_\text{sat} = \frac{p_\text{sat}(T)}{RT}
  \end{equation}
  The rate of evaporation and condensation is given as follows.
  \begin{equation}
    \gamma_{ec} = \left\lbrace \begin{matrix}
    \gamma_e(T)s_\text{red}, & c_{\text{H}_\text{2}\text{O}} < c_\text{sat} \\
    \gamma_c(T)(1-s_\text{red}) & c_{\text{H}_\text{2}\text{O}} > c_\text{sat}
    \end{matrix} \right.
  \end{equation}
  Here, $s_\text{red}$ is the reduced liquid water saturation and is given as $s_\text{red} = (s - s_\text{im})/(1-s_\text{im})$ with $s_{im}$ referring to the immobile saturation.

   \section{Impact of a smaller training dataset}
  
  \begin{figure}[!h]
      \centering
      \includegraphics[width=1.0\textwidth]{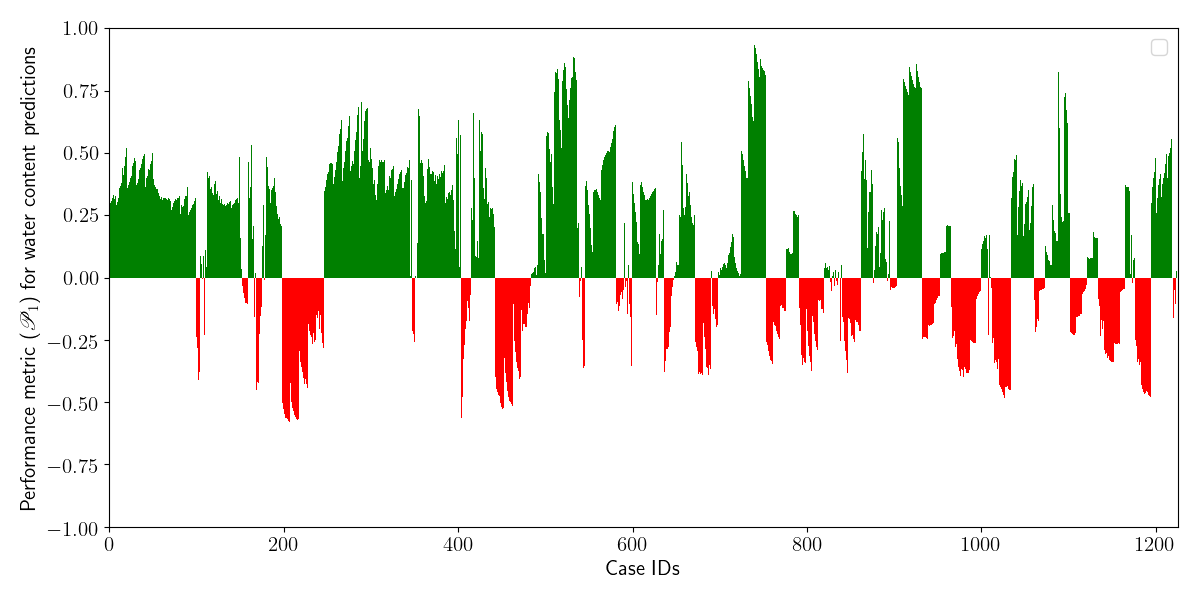}
      \caption{Performance metric for all cases when trained with only 7 (instead of 14) cases}
      \label{fig:performance_water_sparseTraining}
  \end{figure}
  To illustrate the impact of removing training configurations, a second model was trained using only 7 training configurations (case IDs 40, 125, 190, 400, 740, 865 and 1090) instead of 14, and as can be seen from Fig. \ref{fig:performance_water_sparseTraining}, the performance of the augmented model immediately deteriorates and it is able to achieve better-than-baseline performance for only 777 cases out of a total of 1224 that it was tested for.
  It can also be seen that the performance metric for many cases deteriorates drastically, while  it improves for a handful of cases.
  These are either cases which were used during training or those that share very similar inflow conditions with the training cases.
  This behavior is caused by the inference and learning process overfitting to the augmentation behavior specific to the few training cases it has been provided with.
  While it improves predictions on the training cases, overfitting is an undesirable outcome as it results in poorer predictive accuracy for cases different than those in the training dataset and hence, hurts generalizability.